\documentclass{article}
\usepackage[utf8]{inputenc}
\usepackage{fullpage}
\usepackage{authblk}
\usepackage{amsfonts}
\usepackage{amssymb}
\usepackage[pdftex,hidelinks]{hyperref}
\usepackage{amsmath}
\usepackage{color}
\usepackage[sort&compress,numbers]{natbib}
\usepackage{graphicx}
\usepackage{wrapfig}
\usepackage{array}
\usepackage[table]{xcolor}
\usepackage{mathtools}
\usepackage{tikz}
\usepackage[normalem]{ulem} 
\usepackage{caption}
\usepackage{subcaption}

\newcommand{\R}[0]{\mathbb{R}}

\title{Transferring Chemical and Energetic Knowledge Between Molecular Systems With Machine Learning}

\author[1]{Sajjad Heydari\thanks{heydaris@myumanitoba.ca}\thanks{These authors contributed equally to this work}}
\author[2]{Stefano Raniolo\thanks{stefano.raniolo@usi.ch}\thanks{These authors contributed equally to this work}}
\author[1,3]{Lorenzo Livi\thanks{lorenzo.livi@umanitoba.ca}\thanks{Corresponding author}}
\author[2,4]{Vittorio Limongelli\thanks{vittoriolimongelli@gmail.com}\thanks{Corresponding author}}
\affil[1]{Department of Computer Science, University of Manitoba, Winnipeg, MB R3T 2N2, Canada}
\affil[2]{Faculty of Biomedical Sciences, Euler Institute, Università della Svizzera italiana (USI), via G. Buffi 13, CH-6900 Lugano, Switzerland}
\affil[3]{Department of Computer Science, University of Exeter, Exeter EX4 4QF, UK}
\affil[4]{Department of Pharmacy, University of Naples “Federico II”, via D. Montesano 49, I-80131 Naples, Italy}

\begin{document}

\maketitle

\begin{abstract}
Predicting structural and energetic properties of a molecular system is one of the fundamental tasks in molecular simulations, and it has applications in chemistry, biology, and medicine. In the past decade, the advent of machine learning algorithms had an impact on molecular simulations for various tasks, including property prediction of atomistic systems. In this paper, we propose a novel methodology for transferring knowledge obtained from simple molecular systems to a more complex one, endowed with a significantly larger number of atoms and degrees of freedom. In particular, we focus on the classification of high and low free-energy conformations. Our approach relies on utilizing (i) a novel hypergraph representation of molecules, encoding all relevant information for characterizing multi-atom interactions for a given conformation, and (ii) novel message passing and pooling layers for processing and making free-energy predictions on such hypergraph-structured data. Despite the complexity of the problem, our results show a remarkable Area Under the Curve of 0.92 for transfer learning from tri-alanine to the deca-alanine system. Moreover, we show that the same transfer learning approach can also be used in an unsupervised way to group chemically related secondary structures of deca-alanine in clusters having similar free-energy values. Our study represents a proof of concept that reliable transfer learning models for molecular systems can be designed, paving the way to unexplored routes in prediction of structural and energetic properties of biologically relevant systems.
\end{abstract}


\section{Introduction}

Molecular simulations are nowadays a fundamental field of investigation in applied sciences, from chemistry to biology and medicine \cite{joshi2021,palmer2021,shukla2021,shahbabaei2022}. They are typically used to predict the properties of a system with relatively good accuracy. In the era of artificial intelligence and machine learning (ML), new challenges are posed in this field, trying to exploit the ability of ML algorithms to deal with a large amount of data and extrapolate important, yet not immediately apparent information for the system under investigation. ML techniques have been indeed applied to chemo-informatics problems -- prediction of compounds properties like solubility, toxicity etc. -- thanks to the relative abundance of experimental data \cite{Agostini2012237,ecoli_graph_complexity}. In the last decade, first attempts to employ ML in molecular simulations have also appeared. In particular, ML has been used to predict atomistic properties in molecular systems \cite{jin2019multi,lamim2018toward,noe2020machine,miller2020relevance,butler2018machine}, also using first principle calculations (i.e., quantum mechanics) \cite{hong2021,lee2021,burkle2021}, and, more recently, in identification of free-energy states and slow degrees of freedom in molecular systems \cite{doi:10.1146/annurev-physchem-042018-052331,mccarty2017}.
However, the wealth of data represents a major limitation for ML applications and despite the increasing computing power, the sampling capability of a system's phase space still represents a hindering factor in all ML applications to molecular simulations. The sampling issue is even more evident in biologically relevant macromolecules made by thousands of atoms, like DNA and protein systems. In fact, despite a solid theory based on statistical mechanics \cite{PIETRUCCI201732}, the large size of real molecules and the long timescale of the events under consideration, impede even the most advanced simulation techniques to study macromolecules in realistic conditions. A clear example is drug discovery, where the drug in vivo efficacy is determined by ligand-target binding kinetics (quantified as drug residence time \cite{tonge2017drug,schuetz2017kinetics,tiwary2015kinetics,copeland2016drug}), which is hardly predictable by current simulation methods \cite{limongelliligand}. In fact, the free-energy landscapes of drug-protein interaction are typically characterized by a number of high barriers that separate various metastable states, trapping the simulation in limited parts of the energy landscape for extended periods of time \cite{valsson2016enhancing}. Developing enhanced sampling techniques and coarse-grained representations \cite{kmiecik2016,singh2019,limongelliligand,bernardi2015,raniolo2020,lelimousin2016} has significantly ameliorated the sampling capability. However, that remains insufficient in most of the real cases, characterized by complex, long timescale evolution of the system. As a result, the identification of the most probable, fundamental free-energy states is not feasible.

\begin{figure}[htp!]
    \centering
    \includegraphics[keepaspectratio=true,scale=0.8]{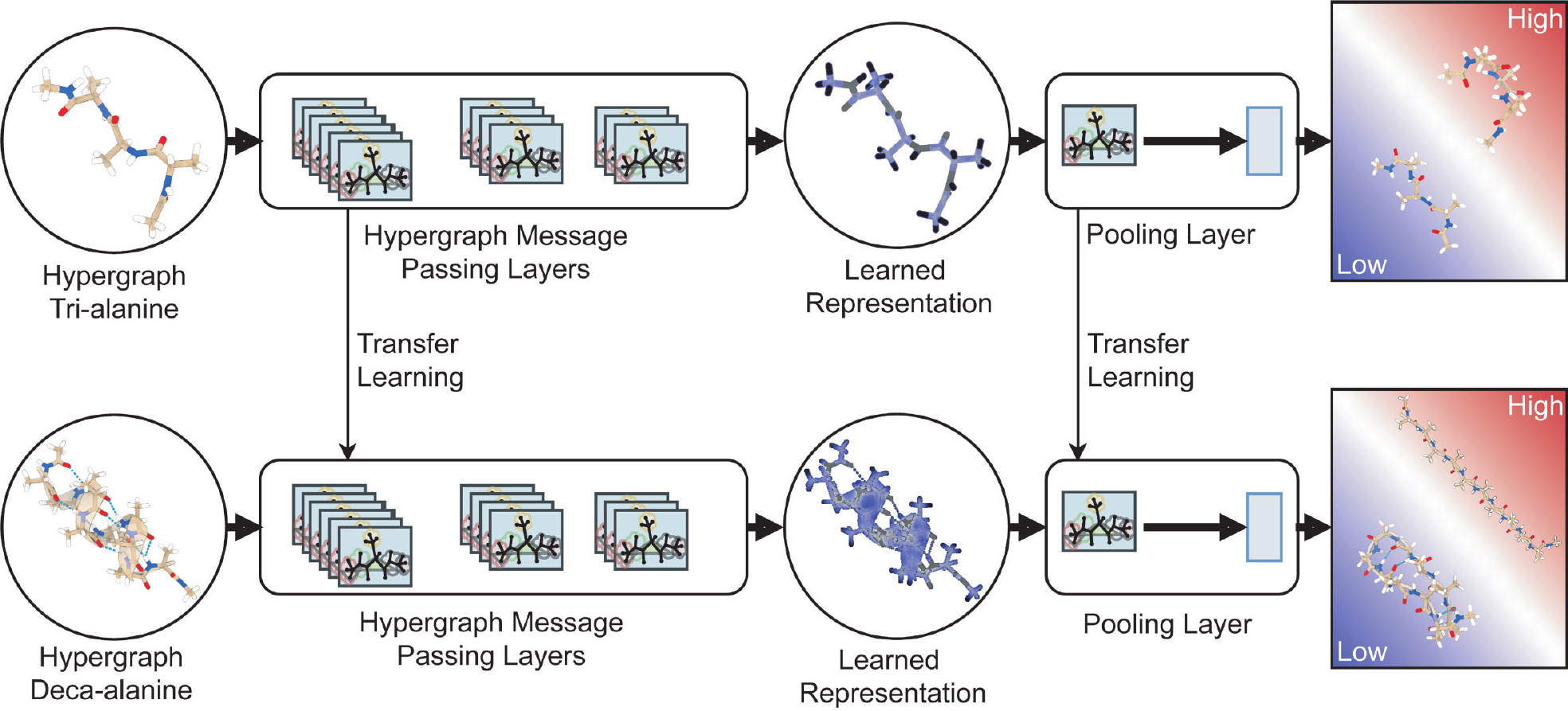}
    \caption{Transfer learning pipeline. The top part of the figure represents the training of the neural network model, where the hypergraph representation of the molecules used for training (e.g. examples of the tri-alanine system) are passed through hypergraph message passing layers to obtain hidden representations. Such representations are further processed by a pooling layer to output the probability of the input being a low free-energy conformation. The bottom part of the figure describes the transfer learning process, where the trained model is used to process examples of the target system (e.g. the deca-alanine system) and make predictions accordingly.}
    \label{fig:pipeline}
\end{figure}

In order to overcome such a limitation, an attracting strategy consists of transferring the knowledge acquired on simple, computationally affordable systems to a much more complex one for predicting relevant properties of the complex system. This strategy is known with the name of transfer learning \cite{weiss2016survey,torrey2010transfer}, and represents a rather unexplored field of investigations in molecular simulations so far \cite{smith2019approaching}.
Here, we address this challenge and propose a novel methodology based on transfer learning that allows learning the free-energy of a given molecular system -- i.e., accurate free-energy data obtained from atomistic simulations -- and transfer such information on a previously unseen molecular system of different size having a significantly larger number of atoms and degrees of freedom that cannot be easily characterized by the free-energy calculations. In particular, we aimed at the classification of low and high free-energy conformations. As shown in Figure \ref{fig:pipeline}, the proposed methodology is based on a novel hypergraph representation of molecules introduced here, which allows encoding all the relevant information for characterizing the multi-atom interactions in a given conformation. The free-energy is then predicted by a novel neural network model capable of processing such hypergraphs as inputs.
Although the literature already contains a few methods based on neural networks for processing hypergraphs \cite{bai2021hypergraph,xia2021self,feng2019hypergraph,jiang2019dynamic} and simplicial complexes \cite{pmlr-v139-bodnar21a}, such methods have some restrictions, e.g. they assume scalars as features for hyperedges and do not offer pooling mechanisms for variable-size inputs, and therefore they are not suitable for the hypergraph representation of molecules introduced here.

We demonstrate the ability of the proposed hypergraph neural network (HNN) on a set of transfer learning experiments. The first one is performed on alanine dipeptide, with the aim to make predictions on the free-energy of a slightly more complex system given by the composition of three alanine peptides, called tri-alanine. Then, we move to a more challenging setting where transfer learning is performed between relatively simple systems (i.e. alanine dipeptide and tri-alanine) and a composition of ten alanine structures, called deca-alanine. This experimental setting represents a real case study since deca-alanine assumes secondary structures which are not present neither in alanine nor in tri-alanine. That is, the most probable conformations of the system, expressed as $\phi$ and $\psi$ torsion angles of alanine, are different in deca-alanine with respect to those assumed in alanine dipeptide and tri-alanine.
Here, we show a remarkable classification performance, quantified by an Area Under the Curve (AUC) of 0.92.
We also show that the same transfer learning approach can be used in an unsupervised way to group chemically related secondary structures of deca-alanine in clusters having similar free-energy values.

Our work is a proof of concept that it is possible, by means of a purposely built machine learning model, to predict free-energy values of a complex molecule using free-energy and structural data of a smaller, yet chemically related molecule, thus \textit{de facto} overcoming the sampling issue for large systems.

\section{Results}
\label{sec:results}

\subsection{Molecular representation and processing}
\label{sec:molecular_representation}

The very first challenge in employing ML to study molecular systems is to develop a reliable molecular representation that is amenable to processing via ML algorithms.
Two important properties that are desirable for molecule representations are uniqueness and invertibility \cite{elton2019deep}.
Uniqueness means that each molecular structure is associated with a single representation; invertibility means that each representation is associated with a single molecule, hence giving rise to a one-to-one mapping.
Most representations used for molecular generation are invertible, but some are not unique \cite{sanchez2018inverse,ceriotti2019unsupervised,schmidt2019recent,chen2019graph}.
There are several reasons for non-uniqueness, including the representation not being invariant to the underlying physical symmetries of rotation, translation, and permutation of atomic indexes. While machine learning algorithms may be directly applied on physical 3D coordinates of atoms, it is preferable removing invariances by creating a more compact representation (removing degrees of freedom) and thus developing a unique representation for each molecule based on internal coordinates only.

Moreover, to be effective for the task at hand, the representation needs to encode both the structural and the physico-chemical properties of the system under investigation. Typically, multi-atom interactions are assessed by computing the potential energy $E_p$ of a structure \cite{kukol2015molecular}, which is classically modelled \cite{andrew2001molecular} as the sum of four parts:
\begin{equation}
\label{eq:potential_energy}
E_p = E_{\mathrm{bond}} + E_{\mathrm{non-bond}} + E_{\mathrm{angle}} + E_{\mathrm{dihedral}}
\end{equation}
This implies that $E_p$ \eqref{eq:potential_energy} cannot be described by only accounting for the interaction between pairs of atoms (dependency on bond length, $E_{\mathrm{bond}}$, and electrostatic interactions, $E_{\mathrm{non-bond}}$).
In fact, the potential energy contains terms that account for angles, $E_{\mathrm{angle}}$, and dihedrals $E_{\mathrm{dihedral}}$ (i.e. the angle formed by two planes defined by four atoms), which are determined by considering the interaction of three and four atoms, respectively.
Accordingly, the commonly used graph representations for molecules \cite{ceriotti2019unsupervised} are not able to fully capture the information required to describe the potential energy \eqref{eq:potential_energy}.

Therefore, we decided to represent each molecule conformation as a hypergraph (see Methods for technical details), with vertices $V$ representing atoms and hyperedges $E=\{e_1, e_2, \cdots, e_N\}$ the various types of interactions among them. In hypergraphs, each hyperedge $e$ is a set and hence it is able to describe the relation between possibly many vertices, i.e. more than two vertices. Notably, we consider $|e|=2$ for bonds and non-bonds interactions, Coulomb and Van der Waals forces, $|e|=3$ for angles between three atoms, and finally $|e|=4$ for the dihedrals between planes formed by four atoms.
A hyperedge feature set of size five is chosen, which stores an encoding of the type of interaction and the related feature value (e.g. the Van der Waals force). A vertex feature set of size two is chosen, which includes the mass and radius of the corresponding atom.  In such a way, nodes and hyperedges of the hypergraphs are equipped with numerical features that ensures an accurate description of the interactions between atoms in each conformation assumed by the system.

Once defined an accurate representation of the system, we fed the HNN with conformations of one system (the simplest one) each labelled with a free-energy value computed through metadynamics calculations. In particular, we consider a system's conformation described by a set of coordinates $x$ and a user-defined metadynamics bias $V(x)$ as a function of a limited number of collective variables $s(x)$, which are functions of the coordinates $x$ (see refs. \cite{Laio2002,Barducci2008} and related Supplementary Information for details). The free energy of the system $F(s)$ as a function of $s(x)$ can be computed as:
\begin{equation}
    \label{eq:free_energy}
    F(s)=-\frac{1}{\beta}\ln\left (\int  dx \ \exp(-\beta V(x))\delta(s-s(x)) \right ),
\end{equation}
where $\beta$ is the inverse of the product of the Boltzmann constant $k_b$ and the temperature $T$ of the system, while $\delta(\cdot)$ is the Dirac delta function. The so computed free-energy ($F(s)$) allows identifying the lowest energy, hence the most probable conformations of a system.

As a result, our model embeds the potential energy \eqref{eq:potential_energy} in the proposed hypergraph representation of molecules, and predicts the free-energy \eqref{eq:free_energy} of a given conformation by inputting such representation to a neural network model. More precisely, let us denote with $\mathcal{H}$ the space of all hypergraphs representing all possible molecular conformations of the system under analysis, and let us denote with $\mathcal{F}$ the space representing the free-energy values (typically $\mathcal{F}=\mathbb{R}$ is the real line). The neural network can be described as a non-linear and parametric function $g: \mathcal{H}\rightarrow\mathcal{F}$ that outputs a free-energy value $f\in\mathcal{F}$ given an input $h\in\mathcal{H}$, i.e. $g(h) = f$. As mentioned before, the neural network is trained with free-energy data obtained from metadynamics simulations.

In the transfer learning setting taken into account here, both the molecule representation and the neural network need to manage two differently sized molecular systems. Hypergraphs naturally account for this aspect by considering a variable number of vertices and hyperedges.
However, neural network models capable to make global predictions on variable-size hypergraphs are not available in the literature. Therefore, we designed a novel message passing layer that can process hypergraph-structured data of variable size, and a novel pooling layer to aggregate the information of variable-size conformations (see Methods for details).

The proposed methodology was tested on molecular systems of different complexity, the alanine, tri-alanine, and deca-alanine systems, and the results are described in the following sections.

\subsection{From alanine dipeptide to tri-alanine}
\label{sec:alanine_trialanine}

In the first experiment, we perform transfer learning from alanine dipeptide to tri-alanine. Alanine dipeptide is a relatively simple molecule, used as reference system for conformational free-energy calculations \cite{sultan2018,mori2020,belkacemi2022}.
In particular, it is well-known that the backbone dihedral angles $\phi$ and $\psi$ are the most relevant degrees of freedom and as such they can distinguish the different conformations assumed by the system. In order to test the transfer learning ability of our model from alanine dipeptide to a more complex and biologically relevant system, we decided to study tri-alanine. In fact, tri-alanine represents a natural evolution of alanine dipeptide, however, it increases the complexity of the system with four additional dihedral angles. Although the structure is not long enough to fold in organized secondary structures (i.e., hairpin or helix), the number of possible conformations is considerably higher than alanine dipeptide. These conformations can be distinguished based on the combination of $\phi$ and $\psi$ angles of each residue that, taken singularly, closely reproduce the behaviour seen in alanine dipeptide. For this reason, it should be feasible to train a neural network on the simpler system, trying to predict characteristics that are also relevant for the more complex one.

The structural and energetic data for alanine dipeptide were obtained from 100 ns metadynamics calculations in vacuum, using $\phi$ and $\psi$ as collective variables (more information in the Methods section). The relative simplicity of the system allowed us to reach convergence of the free-energy calculation, thus providing a reliable ground truth for the HNN model. Instead, the tri-alanine system required 400 ns of metadynamics simulation with a more complex setup that is described in the Methods section.
The conformations and the free-energy data generated by metadynamics represent the input of the HNN model, which is formed by two consecutive layers of message passing followed by a pooling layer, and a single linear layer that outputs the probability of the input being a low free-energy conformation. In particular, the HNN model was trained on the alanine structural and free-energy data, and then the HNN model was transferred to the tri-alanine dataset without using any free-energy information related to the tri-alanine system, i.e. the model was trained in a zero-shot fashion.
The available data is split in training, validation, and test datasets.
For any 5 consecutive conformations in the alanine dataset, the first one was selected for training (20\%), the second and third for validation (40\%), and the fourth and fifth for testing (40\%).

\begin{figure}[h!]
    \centering
    \includegraphics[width=0.70\textwidth, angle=270]{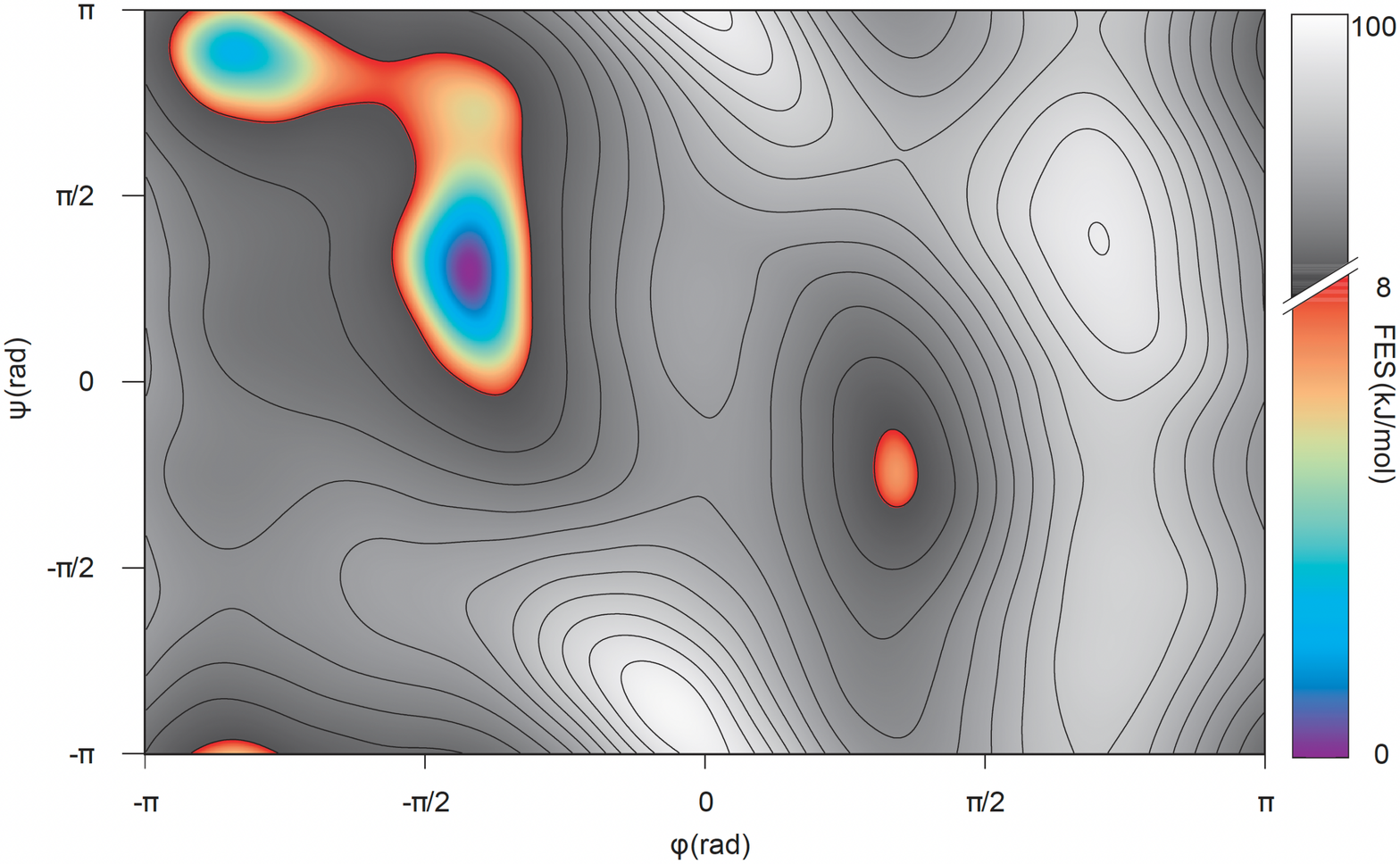}
    \caption{Free-energy surface for the conformational space of alanine dipeptide. In the colour range the interval from 0 to 8 kJ/mol employed for thee threshold. The rest of the surface represents high energy conformations that have been greyed out.}
    \label{fig:8kj}
\end{figure}

As said before, we are interested in evaluating whether HNN can distinguish between high and low free-energy conformations of the tri-alanine system. To this end, we set a threshold value equals to 8 kJ/mol for differentiating between high and low free-energy conformations. This value was chosen considering that all structures comprised in the 0-8 kJ/mol interval, where 0 kJ/mol is attributed to the global minimum, belong to known low energy and metastable states for alanine dipeptide, whereas values greater than 8 kJ/mol correspond to high energy conformations (Fig. \ref{fig:8kj}).

\begin{figure}
    \centering
    \begin{subfigure}[b]{0.25\textwidth}
      \centering
      \includegraphics[width=0.8\textwidth]{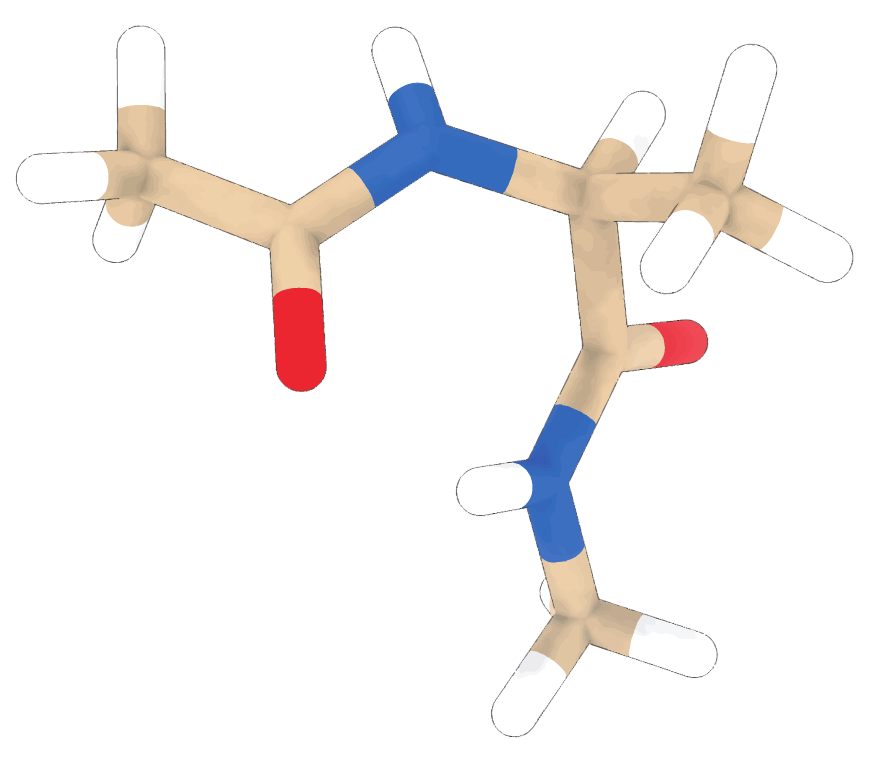}
      \caption{A sample conformation of alanine dipeptide.}
    \end{subfigure}
    \hfill
    \begin{subfigure}[b]{0.25\textwidth}
      \centering
      \includegraphics[width=0.8\textwidth]{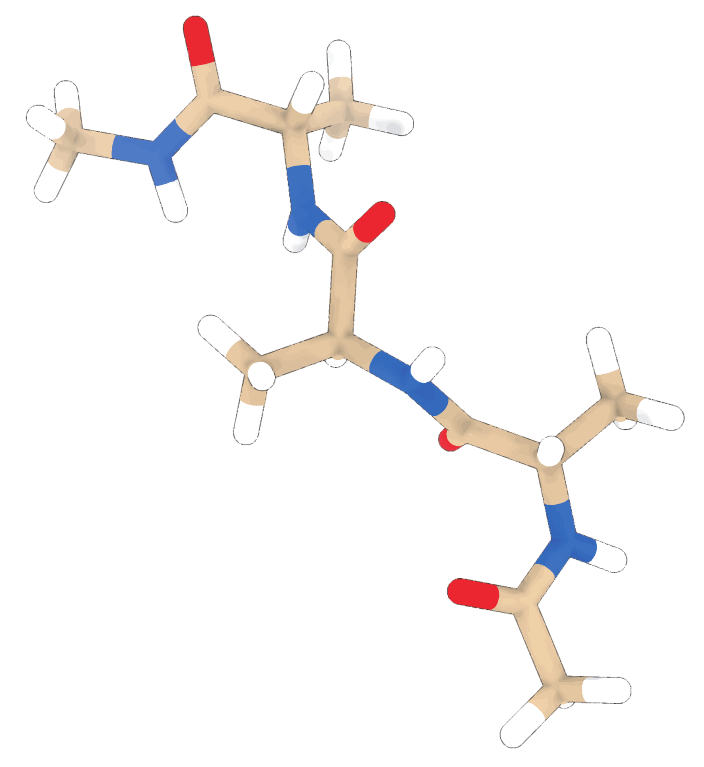}
      \caption{A sample conformation of tri-alanine.}
    \end{subfigure}
    \hfill
    \begin{subfigure}[b]{0.45\textwidth}
        \centering
        \includegraphics[width=0.82\textwidth]{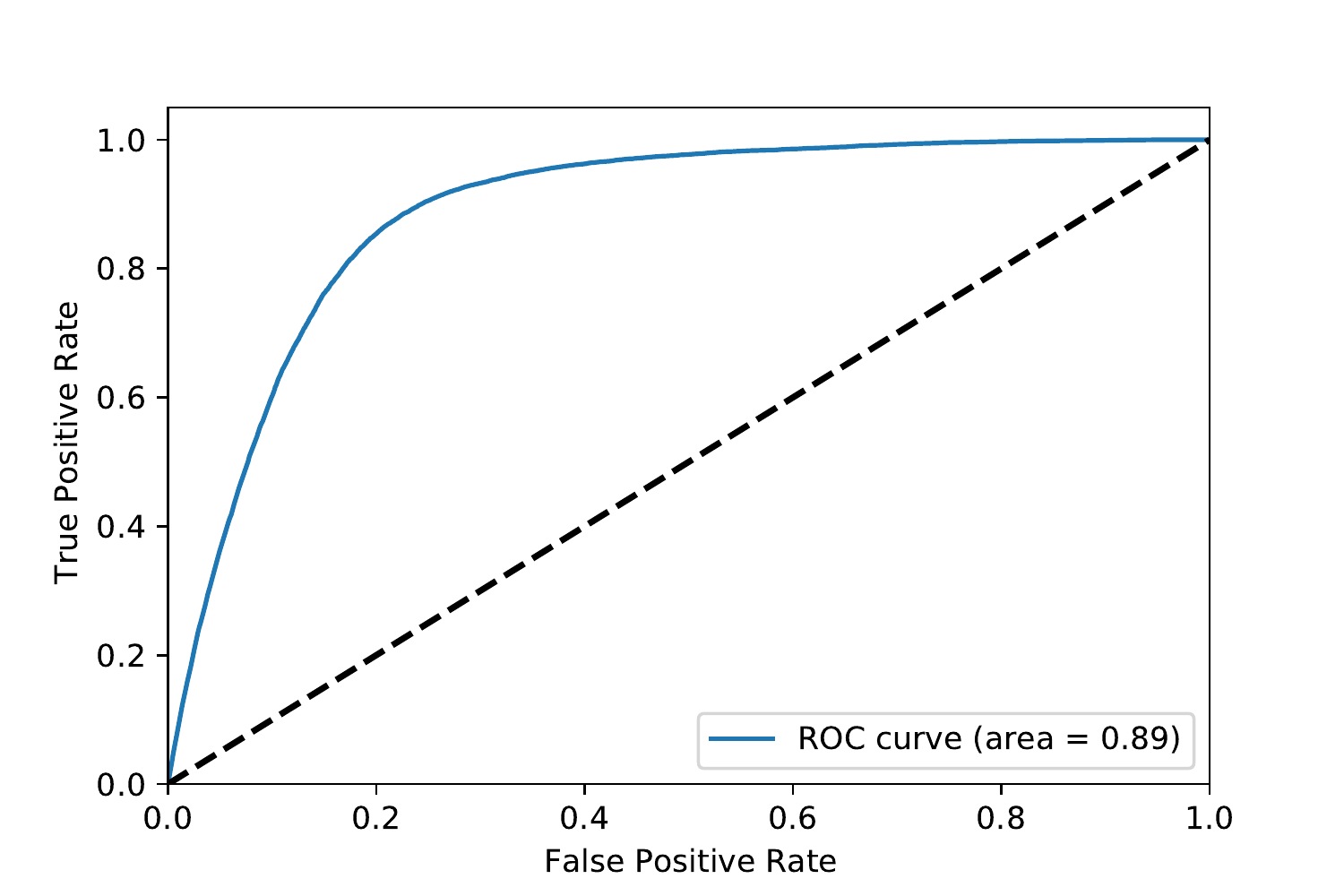}
        \caption{ROC and AUC (0.89) of the HNN model trained on alanine dipeptide and tested on tri-alanine.}
        \label{fig:tri_roc_curve}
    \end{subfigure}
    \caption{Molecules (a, b) used in training and testing of the HNN model along with its ROC and AUC (c).}
\end{figure}

HNN outputs a probability value $p\in[0, 1]$ that the input denotes a low energy conformation of the target system. This output is converted into a deterministic decision by setting a threshold $t\in[0, 1]$: $p\geq t$ indicates membership to the low energy class; conversely, $p<t$ indicates membership to the high energy class. In order to systematically evaluate the performance of our HNN model, and make the performance evaluation not dependent on the choice of threshold $t$, we performed the Receiver Operating Curve (ROC) analysis \cite{Fawcett_ROC}.
The resulting ROC curve is shown in Figure~\ref{fig:tri_roc_curve}, denoting a relatively high AUC value of 0.89.

\subsection{From tri-alanine to deca-alanine}
\label{sec:trialanine_decaalanine}

\begin{figure}[ht]
    \centering
    \includegraphics[scale=0.80]{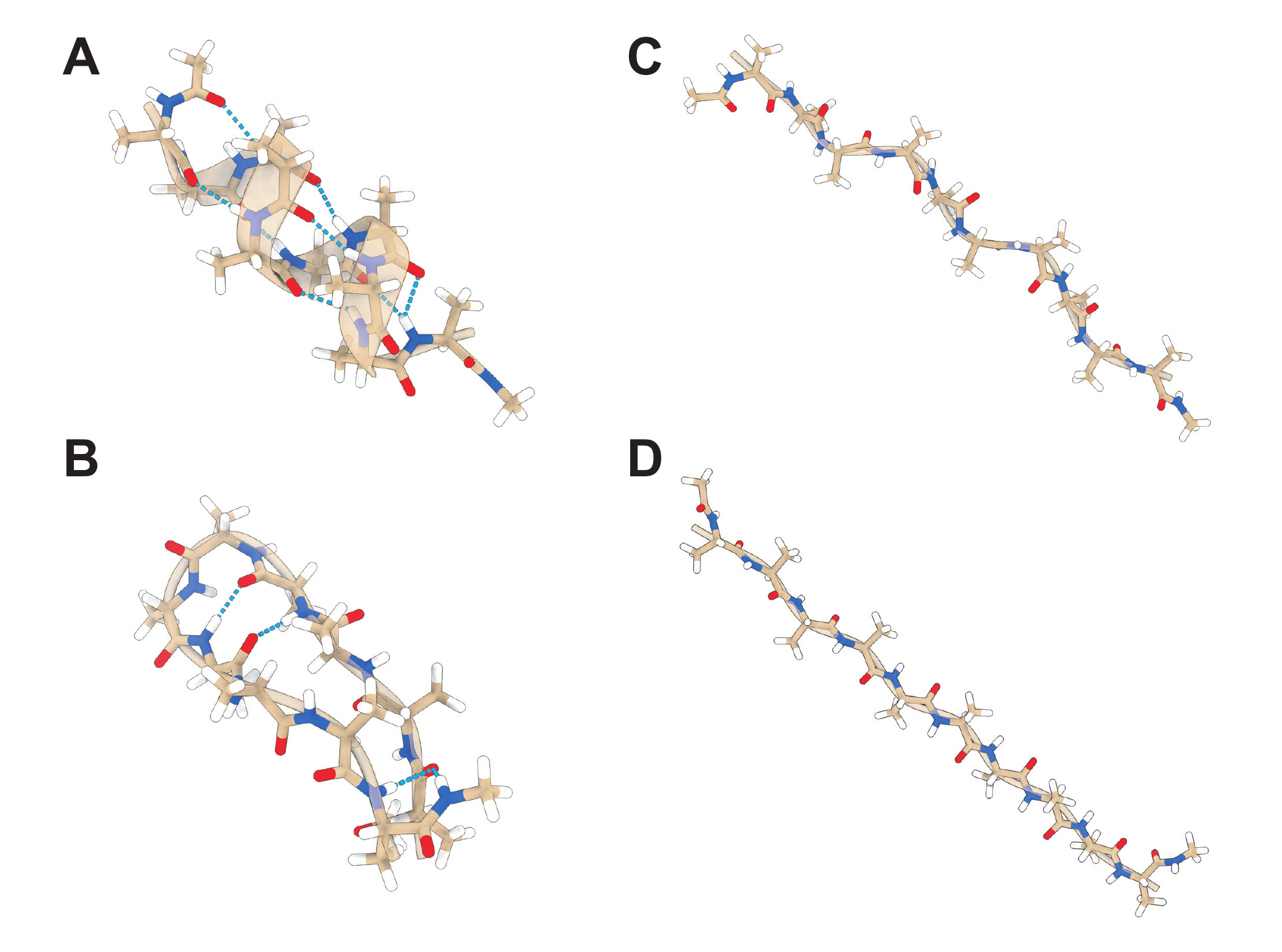}
    \caption{Deca-alanine in $\alpha$-helix (A), hairpin (B), poly-proline (C), and unfolded (D) conformation. Atoms are colored by element (i.e., hydrogen in white, carbon in tan, nitrogen in blue, and oxygen in red), the secondary structure is represented by transparent ribbons, and h-bonds are showed in cyan, whenever they are relevant.}
    \label{deca_conformations}
\end{figure}

The second, biologically more relevant case study considers transfer learning from tri-alanine to deca-alanine. In fact, among the poly-alanine peptides, deca-alanine represents a challenging molecule since it is able to assume secondary structures, characterized by specific alanine conformations that are not represented in alanine dipeptide and tri-alanine systems. The significantly higher structural complexity also increases the difficulty of predicting the free-energy. In Figure \ref{deca_conformations}, we report a selection of possible structures assumed by deca-alanine in vacuum.

The deca-alanine system has been employed as reference model by several groups in order to rank energetics in peptide folding and to test new sampling methods \cite{ozer2012,ozer2014,kokubo2011,post2019,chen2022}. Previous works agree in reporting as the energetically preferred state the helical conformation, passing to higher energy conformations from $\alpha$-helix, to $\pi$-helix, and finally to random coil for the unfolding state \cite{ozer2012}. Alternative structures can also be found (i.e., $\beta$ hairpin), showing proportional or even lower free-energy estimates with respect to the helixes family \cite{post2019}, thus making deca-alanine a real case study of practical importance.

The deca-alanine system was simulated in vacuum for around 700ns. Similarly to what has been done in the tri-alanine case, the sampling of all possible secondary structures was obtained by enhancing the sampling through metadynamics, using the Root Mean-Squared Deviation (RMSD) of the C$_{\alpha}$ atoms of each alanine residue as the Collective Variable (CV) (more information is provided in the Supporting Information). We note that in this case we resorted to metadynamics merely to generate very different conformations of the system by enhancing the sampling of the phase space, while we were not interested in computing the free-energy (see Fig. SI~\ref{fes}). To this end, more sophisticated simulation settings \cite{bonati2021} might be used to take into account the most relevant slow degrees of freedom of the system, though requiring a long and non-trivial procedure.
We note that both deca-alanine and tri-alanine are made by the same building block (i.e. alanine). However, the behaviour of the deca-alanine system is completely different with respect to that of tri-alanine, as the former is able to engage intra-molecular interactions that stabilize specific secondary structures.

\subsubsection{Classification of low and high free-energy conformations}

\begin{figure}[h!]
    \centering
    \includegraphics[width=0.82\textwidth]{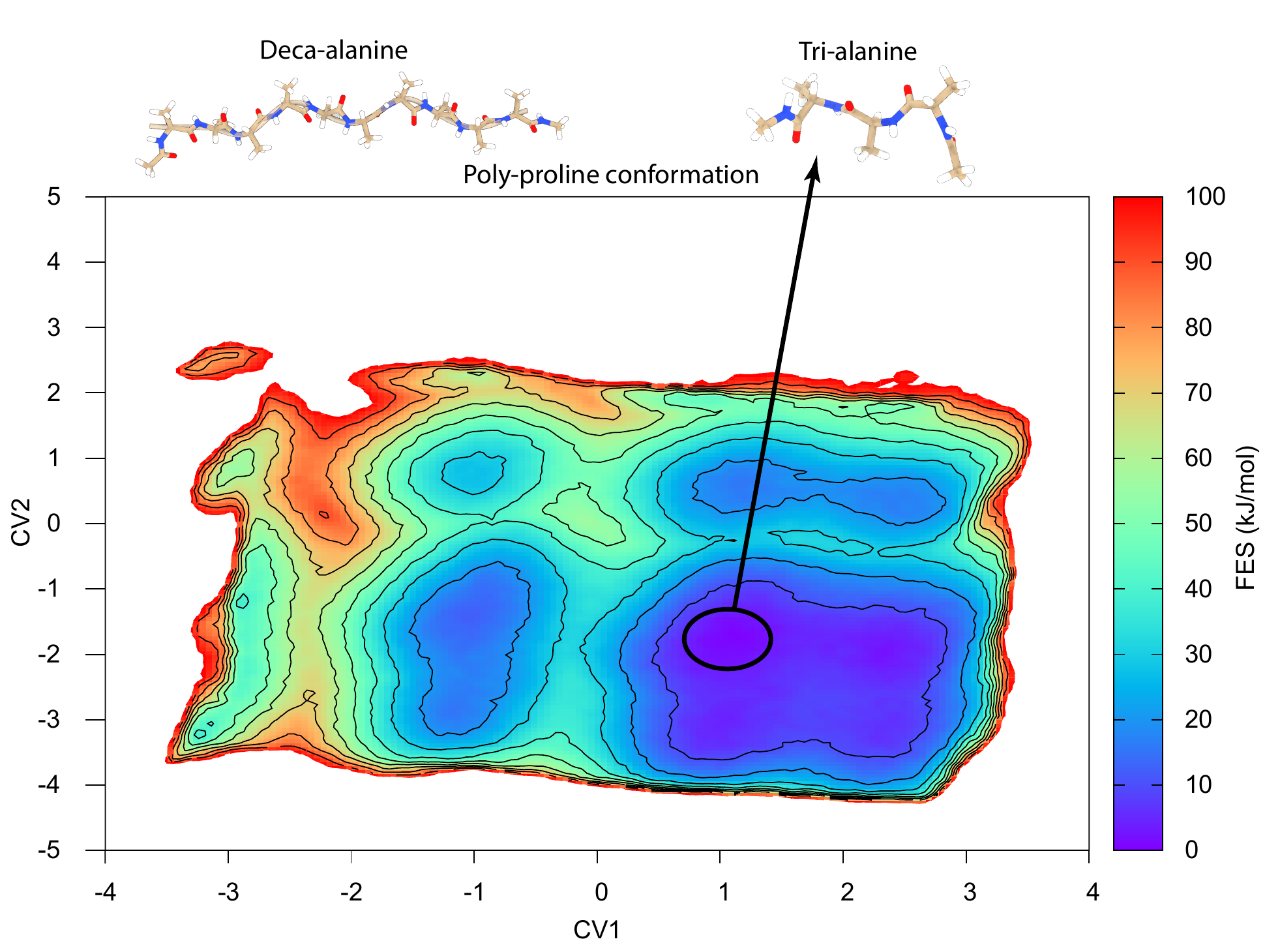}
    \caption{Free-energy surface for tri-alanine. The absolute minimum in the plot corresponds to the poly-proline-like conformation, which is not the preferred conformation in the case of deca-alanine, represented in the upper left corner.}
    \label{fig:triala_fes}
\end{figure}

Here, we assessed the feasibility of training our HNN on tri-alanine structures and the corresponding free-energy data, transferring the acquired knowledge to classify the deca-alanine conformations as low/high free-energy conformations, again in a zero-shot fashion.
We stress that no free-energy estimate of the deca-alanine system was used as supervised information for training the HNN model.
This classification problem is more difficult than what it might seem. In fact, it is interesting to consider that the poly-proline structure, seen as a minimum for tri-alanine, should be rather disfavoured in deca-alanine, which instead prefers assuming conformations stabilzed by intra-molecular h-bond interactions (Fig. \ref{fig:triala_fes}). 

This kind of interaction is indeed present in any helix and $\beta$-sheet secondary structure. One single h-bond typically brings a weak energetic contribution (0.5-6 kcal/mol) \cite{sheu2003}. However, the formation of more h-bonds in a molecule can stabilize even higher-order conformations, where the gain in enthalpy, thanks to the formation of such interactions, is significantly higher than the loss in entropy due to a more constrained conformation assumed by the system. As a result, the formation of helices is possible only in peptides made by a relatively high number of amino acids where a number of h-bonds can be engaged. Importantly, no intra-molecular h-bond is observed in the training set, thus further challenging the HNN model.

As in the previous experiment, the HNN model consists of two layers of message passing for hypergraphs, followed by a pooling layer, finally passing the resulting internal representation through a single linear layer that outputs the probability that the input represents a low free-energy conformation. This model is trained on the tri-alanine dataset containing 100,000 examples; the split considers 20\% randomly chosen training data, 20\% for validation, and 60\% test data.

\begin{figure}[ht!]
    \centering
    \begin{subfigure}[b]{0.25\textwidth}
      \centering
      \includegraphics[width=0.8\textwidth]{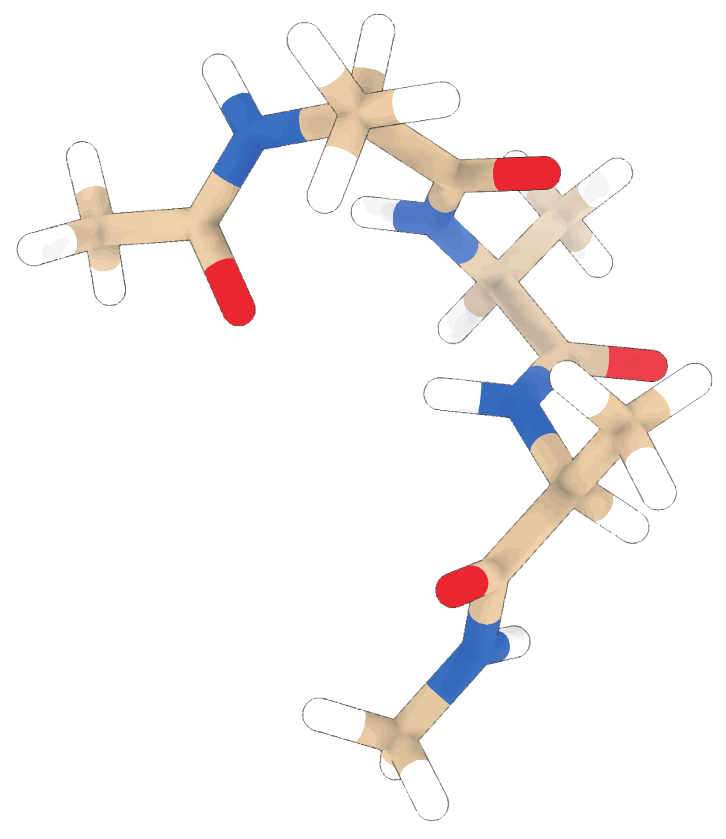}
      \caption{A sample conformation of tri-alanine.}
    \end{subfigure}
    \hfill
    \begin{subfigure}[b]{0.25\textwidth}
      \centering
      \includegraphics[width=1.0\textwidth]{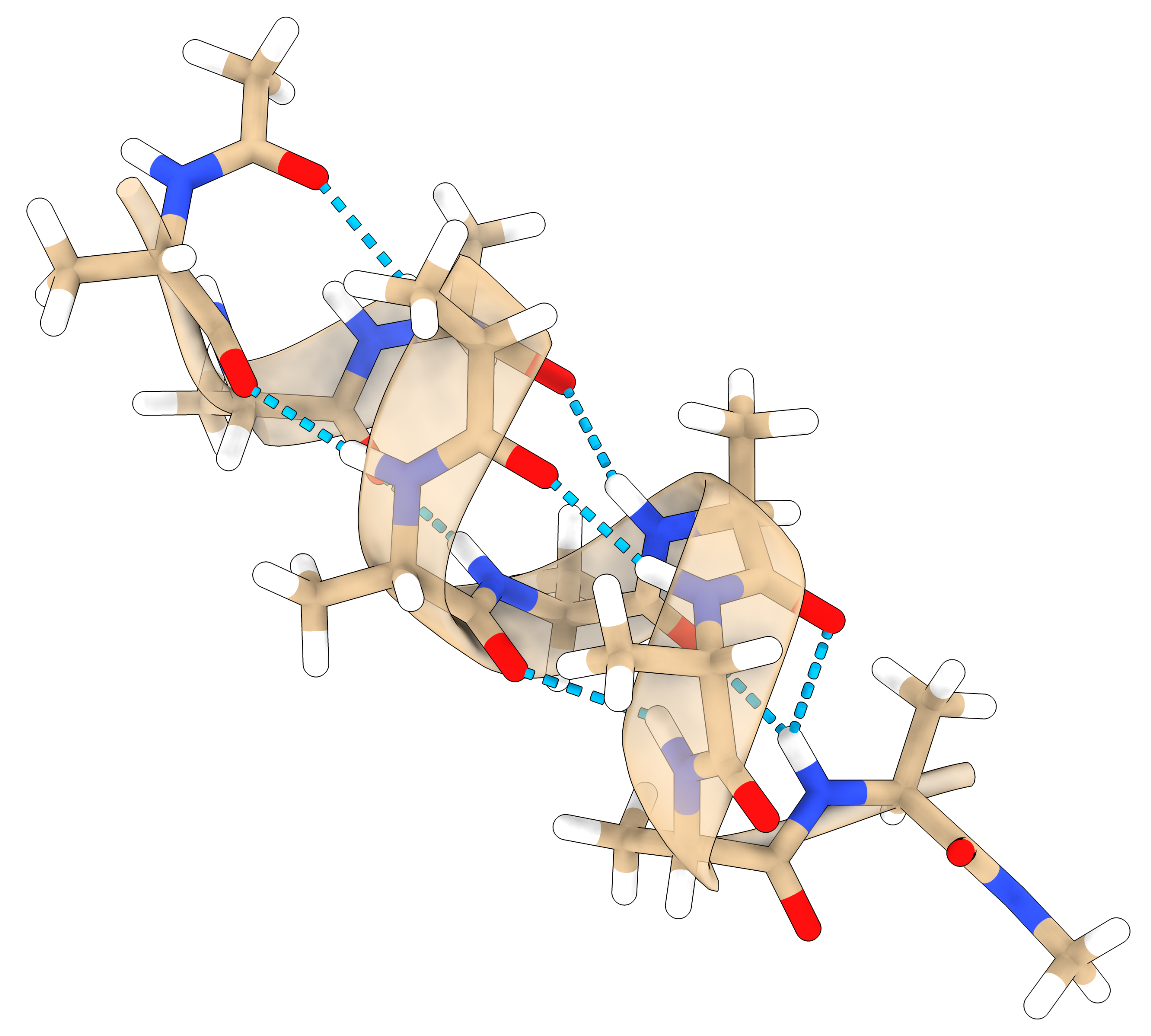}
      \caption{A sample conformation of deca-alanine.}
    \end{subfigure}
    \hfill
    \begin{subfigure}[b]{0.45\textwidth}
        \centering
        \includegraphics[width=\textwidth]{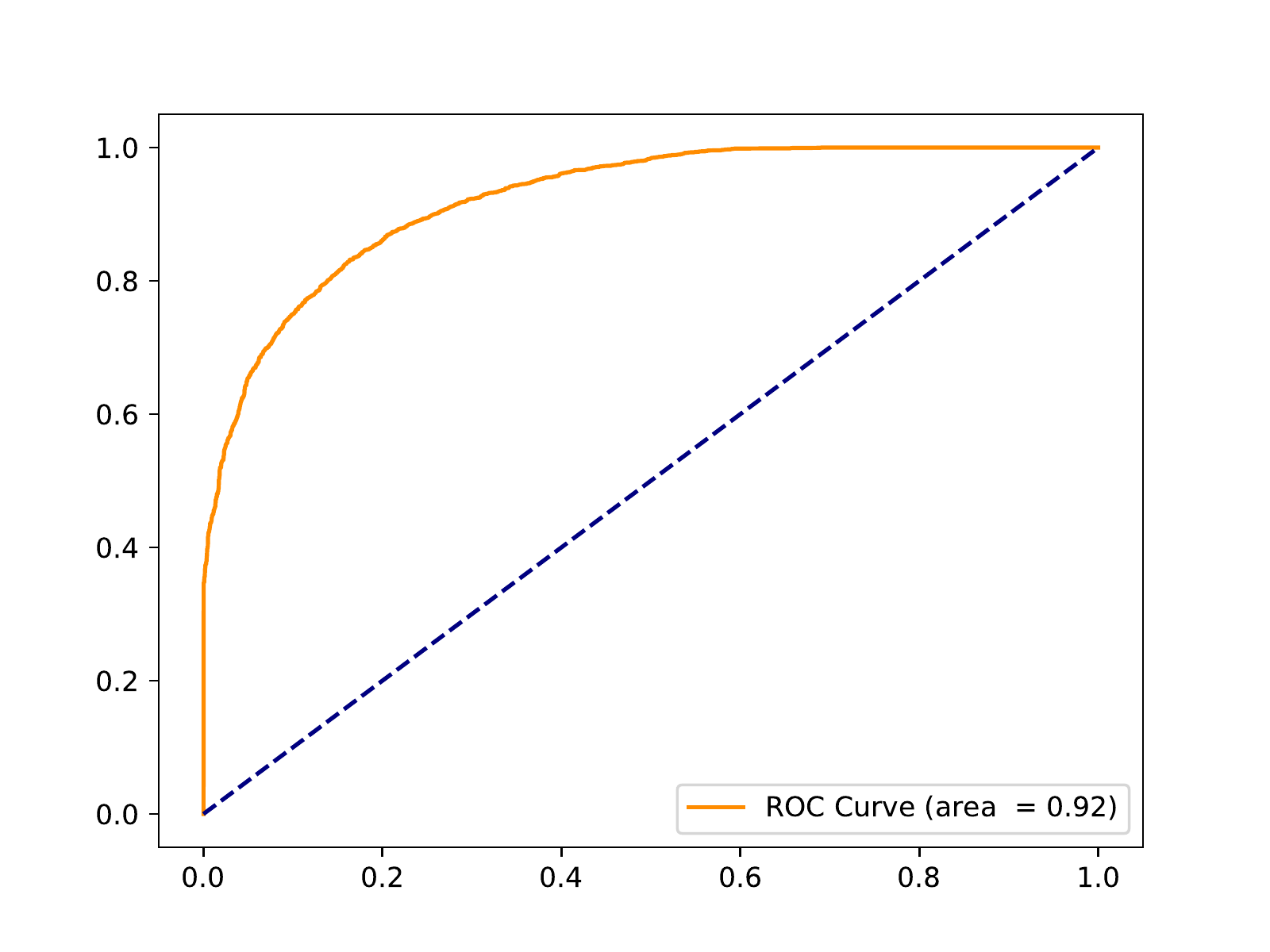}
        \caption{ROC and AUC (0.92) for classification of deca-alanine system.}
        \label{deca-roc}
    \end{subfigure}
    \caption{Molecules (a, b) used in training and testing of the model along with its ROC and AUC (c).}
\end{figure}

The performance of the model over the deca-alanine system varies depending on the chosen threshold for discriminating low and high free-energy conformations. Three different representative values have been selected and results are shown in Table~\ref{tab:deca-prec-rec}.
As in the previous experiment, to provide a more robust measure of classification performance that does not depend on the choice of a specific threshold, we performed ROC analysis and computed the AUC of our HNN model. Results are shown in Figure~\ref{deca-roc}, which denote a remarkable AUC of 0.92, thus confirming the ability of our HNN model to distinguish between high and low free-energy deca-alanine conformations with a remarkable performance.
\begin{table}[htp!]
    \caption{Classification results for deca-alanine with different thresholds.}
    \label{tab:deca-prec-rec}
    \centering
    \begin{tabular}{|c|c|c|}
    \hline
        \textbf{Threshold} & \textbf{Precision} ($\frac{tp}{tp+fp}$) & \textbf{Recall} ($\frac{tp}{tp+fn}$) \\ \hline\hline
        
        Low $\leq$ 0.45 & 0.956 & 0.836 \\\hline
        High$>$0.45 & 0.979 & 0.624 \\\hline \hline
        
        Low $\leq$ 0.5 & 0.923 & 0.950 \\\hline
        High$>$0.5 & 0.921 & 0.880 \\\hline \hline
        
        Low $\leq$ 0.55 & 0.907 & 0.999  \\\hline
        High $>$ 0.55 & 0.877 & 0.894 \\\hline

    \end{tabular}
\end{table}

\subsubsection{Secondary structure recognition}

Obtaining a converged free-energy calculation and the identification of low free-energy states as a ground truth for deca-alanine is not trivial, like for many other complex molecular systems. For this reason, using only the structures generated by the simulations we challenged the HNN model in recognizing different secondary structures in an unsupervised way. More precisely, we used the HNN model to make predictions on the deca-alanine free-energy values and used those predictions to cluster conformations on the sole base of their numerical similarity. Detailed methodological aspects are discussed in Section \ref{sec:secondary_recognition}.

\begin{figure}[ht!]
    \centering
    \includegraphics[scale=0.75]{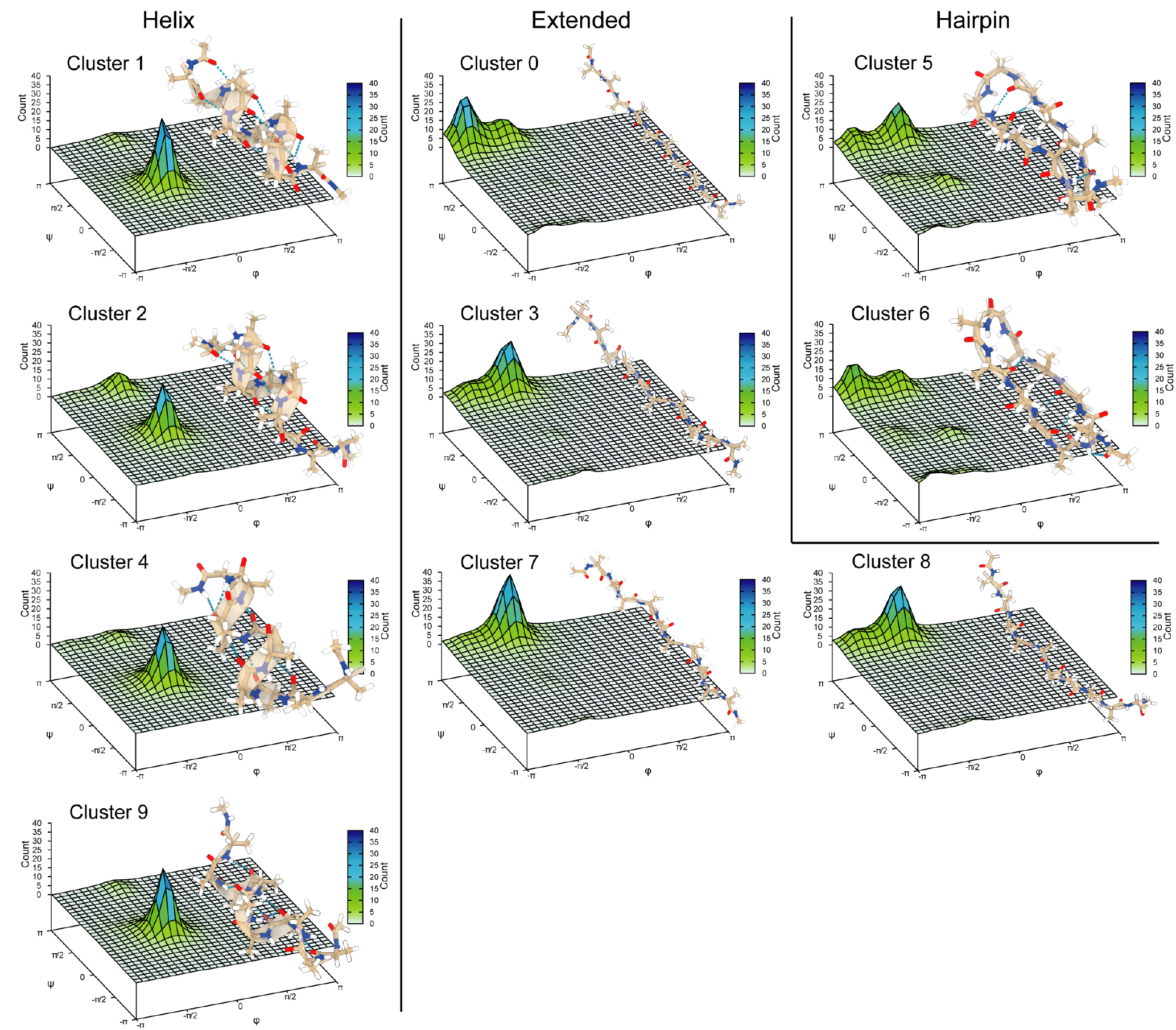}
    \caption{Ramachandran plot for a distribution of 1000 structures per cluster, which have been divided with respect to their family. A structural representation is also offered to show the different conformations of deca-alanine.}
    \label{fig:clusters}
\end{figure}

The deca-alanine conformations generated by the atomistic simulations can be clustered in ten conformational families based on the RMSD of the alanine backbone atoms. Figure \ref{fig:clusters} shows the representation of the ten clusters together with the distribution of their $\phi$ and $\psi$ angles in a 3D Ramachandran plot.
It is important to note that the HNN model does not use the RMSD-based clustering information.
Then, additional structures were generated from the ten most populated clusters by means of standard MD simulations. In particular, each cluster representative has been simulated with a constraint on the RMSD of the backbone atoms to produce 1000 additional structures for each cluster representative, reaching a total of 10000 structures.

The ten different clusters, with numerical identifiers going from 0 to 9, can be grouped in three distinct families, whose members share common structural features that should be recognized by our neural network during transfer learning:
\begin{itemize}
    \item Helix family: clusters 1, 2, 4, and 9;
    \item Hairpin-like family: clusters 5 and 6;
    \item Extended family: all unfolded conformations (i.e., poly-proline and fully extended $\beta$ structures) in clusters 0, 3, 7, and 8.
\end{itemize}

Figure~\ref{fig:p-value-colormap} shows a colour map of the outcome of the statistical tests performed to assess the similarity between the distributions underlying the free-energy predictions made by the HNN model for the structures in the various clusters. 
Green and blue cells denote outcomes that are in agreement with our initial assumptions of energetically dissimilar and similar structures, respectively. On the other hand, yellow and red cells indicate unexpected energetically dissimilar and similar structures, respectively.
In detail, green cells indicate that $p$-value is lower than threshold (0.01) as expected, blue indicates that $p$-value was greater than threshold as expected, yellow indicates that $p$-value was unexpectedly lower than threshold, and red indicates that $p$-value was unexpectedly higher than threshold.
See Methods for technical details on the statistical tests used to assess the differences.

\begin{figure}[ht!]
    \centering
    \includegraphics[scale=0.75]{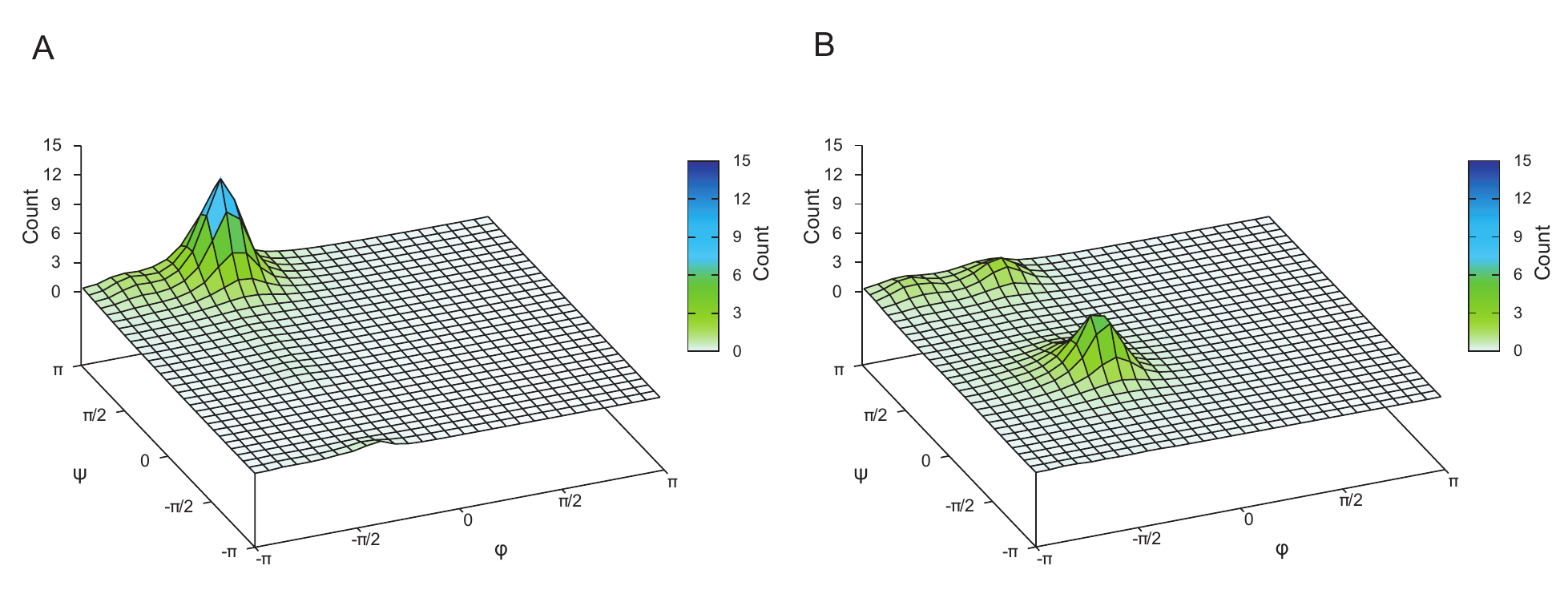}
    \caption{Ramachandran plot for a distribution of 1000 structures of the last three terminal residues in cluster 2 and 4 (A and B, respectively). The different level of denaturation leads to a change in the response of the HNN model.}
    \label{fig:diff}
\end{figure}

Interestingly, the HNN model correctly recognizes structures of diverse clusters that share similar conformational properties, despite some exceptions that are reported in Table \ref{tab:exp-exp}. Among these, the most interesting cases are discussed in the following.
For example, cluster 1, which represents structures with a perfectly folded $\alpha$-helix, is correctly recognized as a low energy conformation, similar to the structures in clusters 4 and 9, but not with respect to cluster 2. The latter is characterized by a helical conformation similar to clusters 1, 4, and 9. However, its  low $p$-values relative to the other clusters indicate that cluster 2 is energetically different from the others. A closer visual inspection of clusters 2 and 4 representative conformations, which are structurally similar, reveals that cluster 2 has the last three residues at C-terminus in a rather unfolded conformation with respect to cluster 4, which has instead an unfolded N-terminus end (see Fig. SI \ref{fig:diff}). Such a minor structural diversity leads to a difference in free-energy that is predicted by the HNN model. Similarly, the HNN model is able to distinguish between clusters 2 and 9, which have minor structural differences as those reported for clusters 2 and 4. On the other hand, cluster 9 and 4, which show a similar secondary structure with the same number of unfolded residues at the same end, are indicated as energetically close by the model.

Another interesting example is cluster 0 with respect to the clusters of the hairpin-like family, i.e., clusters 5 and 6.
The latter is characterized by two $\beta$-sheets organized in an anti-parallel fashion that maximizes the number of intra-molecular h-bonds. On the other hand, cluster 0 is a fully extended $\beta$ structure, with no inter-strand interaction. Despite the similar torsion angles assumed by the alanine residues in 0, 5 and 6, HNN was able to correctly predict the diversity of 0 with respect to 5 and 6, however detecting the similarity between 5 and 6 (see Table~\ref{tab:exp-exp}).

Overall, our HNN correctly predicts most of the energetically and structurally similar conformations, however presenting a number of outliers (e.g., clusters 0 with 3, 0 with 8, 2 with 4, 2 with 9, see Table \ref{tab:exp-exp} for a comprehensive list). Interestingly, for some of them HNN seems sensitive to subtle structural differences between clusters, which would otherwise be considered similar by standard clustering methods such as those based on root mean square deviation (RMSD).

\definecolor{mapColorLThreshold}{rgb}{0,1,0}  
\definecolor{mapColorLThresholdBad}{rgb}{1,1,0} 
\definecolor{mapColorGThreshold}{rgb}{0,0,1} 
\definecolor{mapColorGThresholdBad}{rgb}{1,0,0} 
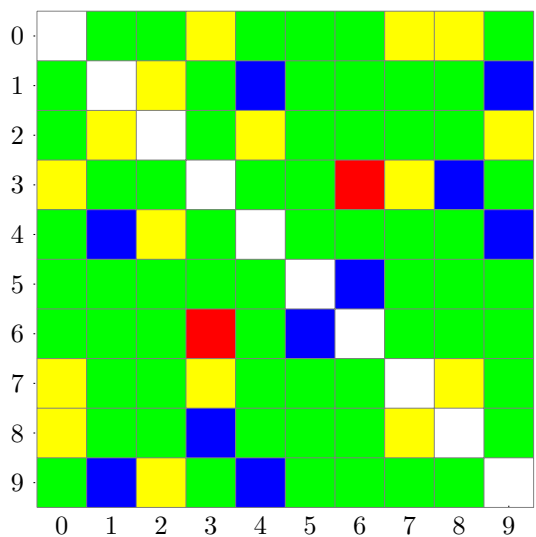
\begin{figure}
    \centering
\begin{tikzpicture}[scale=0.04\textwidth/1cm]
\foreach \x in {0,1,2,3,4,5,6,7,8,9}
  \draw (\x cm+0.5cm, 1pt+1cm) -- (\x cm+0.5cm, 1cm) node[anchor=north] {$\x$};
\foreach \x in {0,1,2,3,4,5,6,7,8,9}
  \draw (-1pt, 10cm-\x cm+0.5cm) -- (-2pt, 10cm-\x cm+0.5cm) node[anchor=east] {$\x$};
\fill[mapColorLThreshold](0,9) rectangle (1, 10);
\fill[mapColorLThreshold](0,8) rectangle (1, 9);
\fill[mapColorLThresholdBad](0,7) rectangle (1, 8);
\fill[mapColorLThreshold](0,6) rectangle (1, 7);
\fill[mapColorLThreshold](0,5) rectangle (1, 6);
\fill[mapColorLThreshold](0,4) rectangle (1, 5);
\fill[mapColorLThresholdBad](0,3) rectangle (1, 4);
\fill[mapColorLThresholdBad](0,2) rectangle (1, 3);
\fill[mapColorLThreshold](0,1) rectangle (1, 2);
\fill[mapColorLThreshold](1,10) rectangle (2, 11);
\fill[mapColorLThresholdBad](1,8) rectangle (2, 9);
\fill[mapColorLThreshold](1,7) rectangle (2, 8);
\fill[mapColorGThreshold](1,6) rectangle (2, 7);
\fill[mapColorLThreshold](1,5) rectangle (2, 6);
\fill[mapColorLThreshold](1,4) rectangle (2, 5);
\fill[mapColorLThreshold](1,3) rectangle (2, 4);
\fill[mapColorLThreshold](1,2) rectangle (2, 3);
\fill[mapColorGThreshold](1,1) rectangle (2, 2);
\fill[mapColorLThreshold](2,10) rectangle (3, 11);
\fill[mapColorLThresholdBad](2,9) rectangle (3, 10);
\fill[mapColorLThreshold](2,7) rectangle (3, 8);
\fill[mapColorLThresholdBad](2,6) rectangle (3, 7);
\fill[mapColorLThreshold](2,5) rectangle (3, 6);
\fill[mapColorLThreshold](2,4) rectangle (3, 5);
\fill[mapColorLThreshold](2,3) rectangle (3, 4);
\fill[mapColorLThreshold](2,2) rectangle (3, 3);
\fill[mapColorLThresholdBad](2,1) rectangle (3, 2);
\fill[mapColorLThresholdBad](3,10) rectangle (4, 11);
\fill[mapColorLThreshold](3,9) rectangle (4, 10);
\fill[mapColorLThreshold](3,8) rectangle (4, 9);
\fill[mapColorLThreshold](3,6) rectangle (4, 7);
\fill[mapColorLThreshold](3,5) rectangle (4, 6);
\fill[mapColorGThresholdBad](3,4) rectangle (4, 5);
\fill[mapColorLThresholdBad](3,3) rectangle (4, 4);
\fill[mapColorGThreshold](3,2) rectangle (4, 3);
\fill[mapColorLThreshold](3,1) rectangle (4, 2);
\fill[mapColorLThreshold](4,10) rectangle (5, 11);
\fill[mapColorGThreshold](4,9) rectangle (5, 10);
\fill[mapColorLThresholdBad](4,8) rectangle (5, 9);
\fill[mapColorLThreshold](4,7) rectangle (5, 8);
\fill[mapColorLThreshold](4,5) rectangle (5, 6);
\fill[mapColorLThreshold](4,4) rectangle (5, 5);
\fill[mapColorLThreshold](4,3) rectangle (5, 4);
\fill[mapColorLThreshold](4,2) rectangle (5, 3);
\fill[mapColorGThreshold](4,1) rectangle (5, 2);
\fill[mapColorLThreshold](5,10) rectangle (6, 11);
\fill[mapColorLThreshold](5,9) rectangle (6, 10);
\fill[mapColorLThreshold](5,8) rectangle (6, 9);
\fill[mapColorLThreshold](5,7) rectangle (6, 8);
\fill[mapColorLThreshold](5,6) rectangle (6, 7);
\fill[mapColorGThreshold](5,4) rectangle (6, 5);
\fill[mapColorLThreshold](5,3) rectangle (6, 4);
\fill[mapColorLThreshold](5,2) rectangle (6, 3);
\fill[mapColorLThreshold](5,1) rectangle (6, 2);
\fill[mapColorLThreshold](6,10) rectangle (7, 11);
\fill[mapColorLThreshold](6,9) rectangle (7, 10);
\fill[mapColorLThreshold](6,8) rectangle (7, 9);
\fill[mapColorGThresholdBad](6,7) rectangle (7, 8);
\fill[mapColorLThreshold](6,6) rectangle (7, 7);
\fill[mapColorGThreshold](6,5) rectangle (7, 6);
\fill[mapColorLThreshold](6,3) rectangle (7, 4);
\fill[mapColorLThreshold](6,2) rectangle (7, 3);
\fill[mapColorLThreshold](6,1) rectangle (7, 2);
\fill[mapColorLThresholdBad](7,10) rectangle (8, 11);
\fill[mapColorLThreshold](7,9) rectangle (8, 10);
\fill[mapColorLThreshold](7,8) rectangle (8, 9);
\fill[mapColorLThresholdBad](7,7) rectangle (8, 8);
\fill[mapColorLThreshold](7,6) rectangle (8, 7);
\fill[mapColorLThreshold](7,5) rectangle (8, 6);
\fill[mapColorLThreshold](7,4) rectangle (8, 5);
\fill[mapColorLThresholdBad](7,2) rectangle (8, 3);
\fill[mapColorLThreshold](7,1) rectangle (8, 2);
\fill[mapColorLThresholdBad](8,10) rectangle (9, 11);
\fill[mapColorLThreshold](8,9) rectangle (9, 10);
\fill[mapColorLThreshold](8,8) rectangle (9, 9);
\fill[mapColorGThreshold](8,7) rectangle (9, 8);
\fill[mapColorLThreshold](8,6) rectangle (9, 7);
\fill[mapColorLThreshold](8,5) rectangle (9, 6);
\fill[mapColorLThreshold](8,4) rectangle (9, 5);
\fill[mapColorLThresholdBad](8,3) rectangle (9, 4);
\fill[mapColorLThreshold](8,1) rectangle (9, 2);
\fill[mapColorLThreshold](9,10) rectangle (10, 11);
\fill[mapColorGThreshold](9,9) rectangle (10, 10);
\fill[mapColorLThresholdBad](9,8) rectangle (10, 9);
\fill[mapColorLThreshold](9,7) rectangle (10, 8);
\fill[mapColorGThreshold](9,6) rectangle (10, 7);
\fill[mapColorLThreshold](9,5) rectangle (10, 6);
\fill[mapColorLThreshold](9,4) rectangle (10, 5);
\fill[mapColorLThreshold](9,3) rectangle (10, 4);
\fill[mapColorLThreshold](9,2) rectangle (10, 3);

\draw[step=1cm,gray,very thin] (0,1) grid (10,11);
\end{tikzpicture}
    \caption{Colour map of $p$-values assessing whether two clusters are in significant agreement in terms of free-energy predictions. Green cells indicate that $p$-value is lower than threshold (0.01) as expected, blue indicates that $p$-value was greater than threshold as expected, yellow indicates that $p$-value was unexpectedly lower than threshold, and red indicates that $p$-value was unexpectedly higher than threshold.}
    \label{fig:p-value-colormap}
\end{figure}

\begin{table}[ht!]
    \caption{Systematic analysis of peculiar results found by the HNN model. The column ``Expectation'' contains the expected outcome based on our visual comparison of the clustered structures, while the ``Explanation'' column gives a justification for the deviation from the expected outcome, highlighting that our model is able to detect subtle differences and similarities between structures in different clusters.}
    \label{tab:exp-exp}
    \centering
    \begin{tabular}{|>{\centering}m{1.5cm}|>{\centering}m{1.5cm}|m{4cm}|m{6cm}|}
        \hline
        \textbf{First cluster} &  \textbf{Second cluster} & \multicolumn{1}{c|}{\textbf{Expectation}} & \multicolumn{1}{c|}{\textbf{Explanation}} \\\hline
        0 \includegraphics[width=0.08\textwidth]{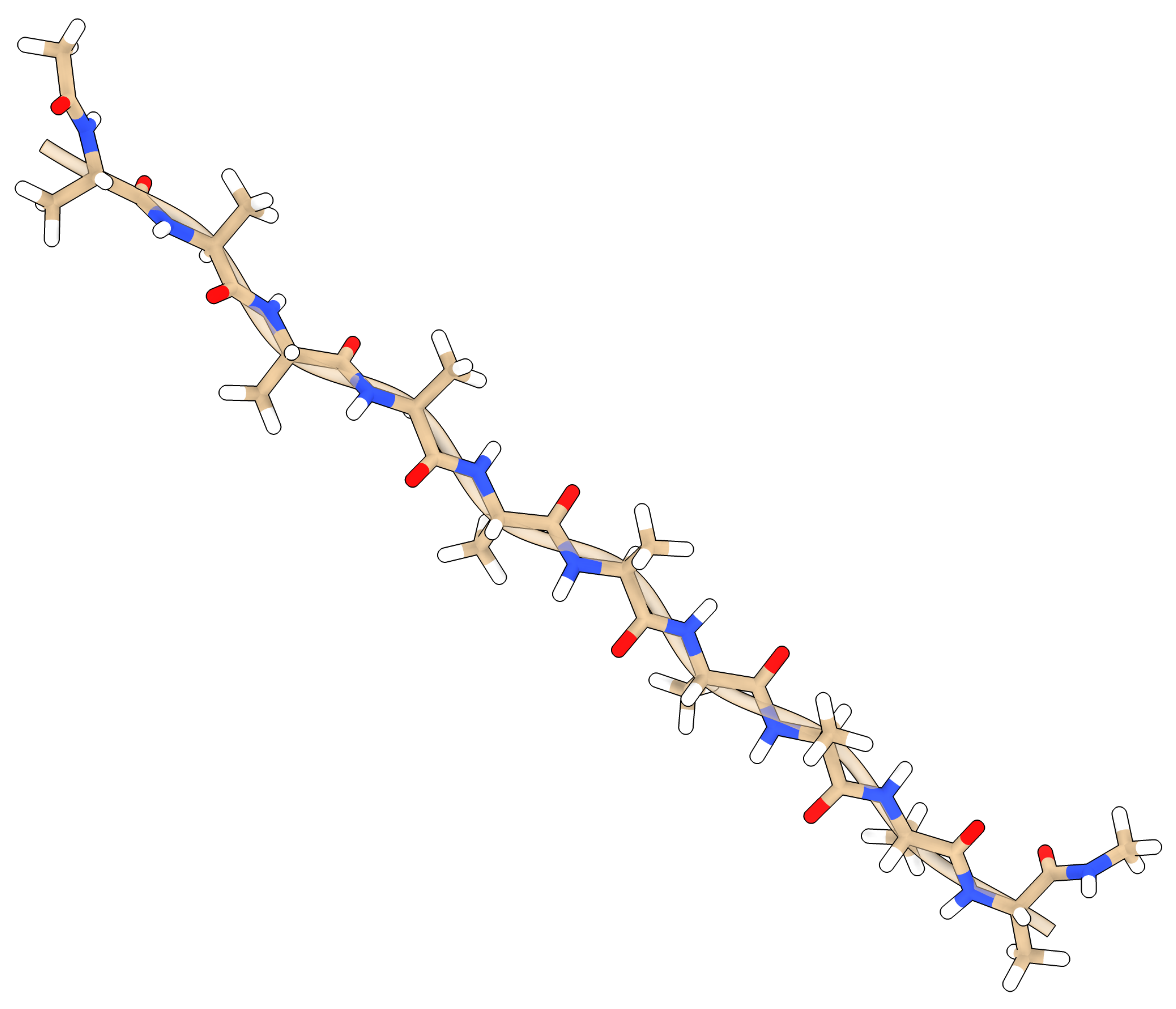} & 3 \includegraphics[width=0.09\textwidth]{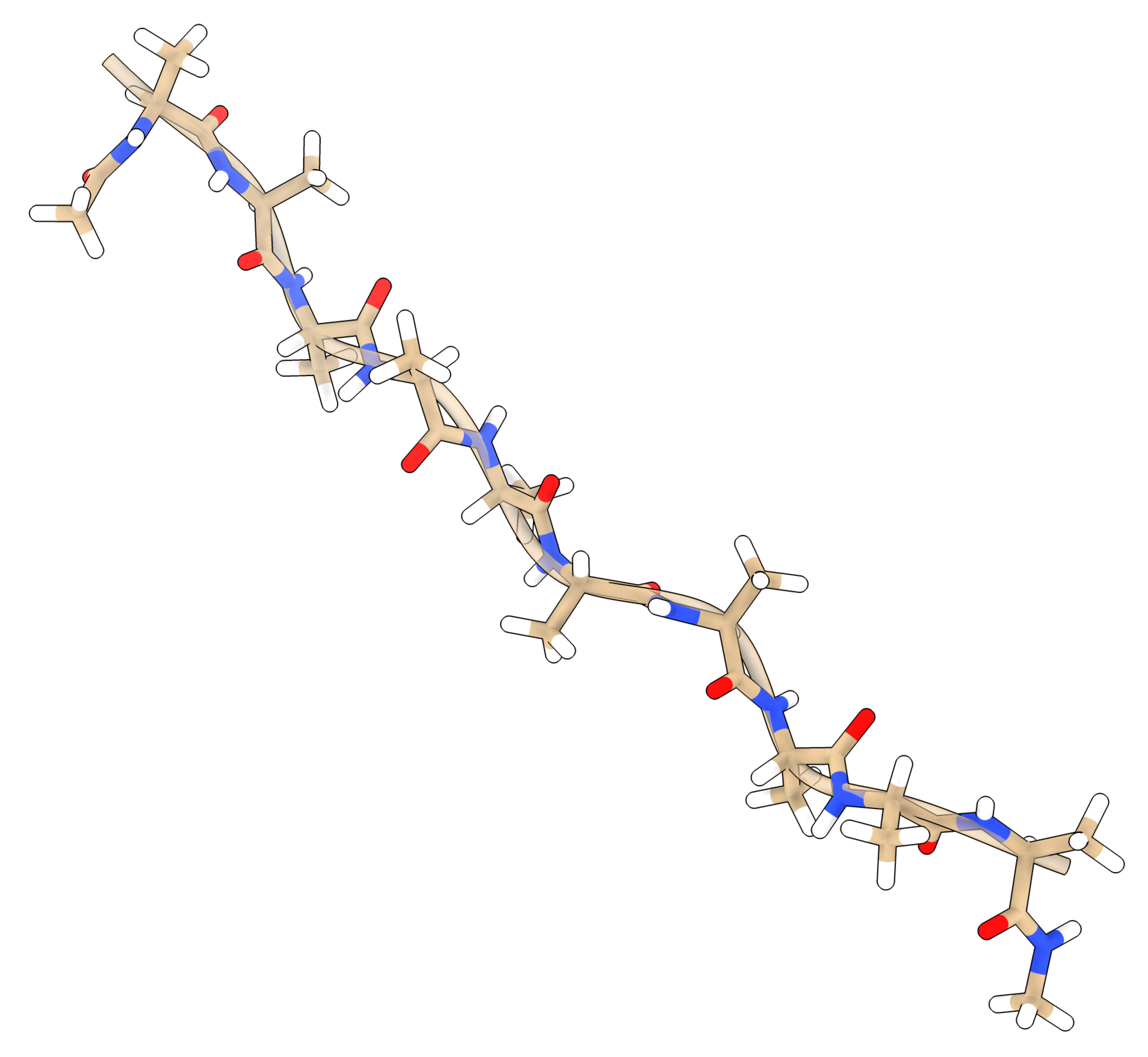} & High similarity for predictions & Cluster 0 belongs to the extended family just like cluster 3, but the latter is mainly organized in poly-proline, thus introducing a significant structural difference that is detected by the HNN model \\
        \hline
        0 \includegraphics[width=0.08\textwidth]{images/deca_beta.pdf} & 7 \includegraphics[width=0.09\textwidth]{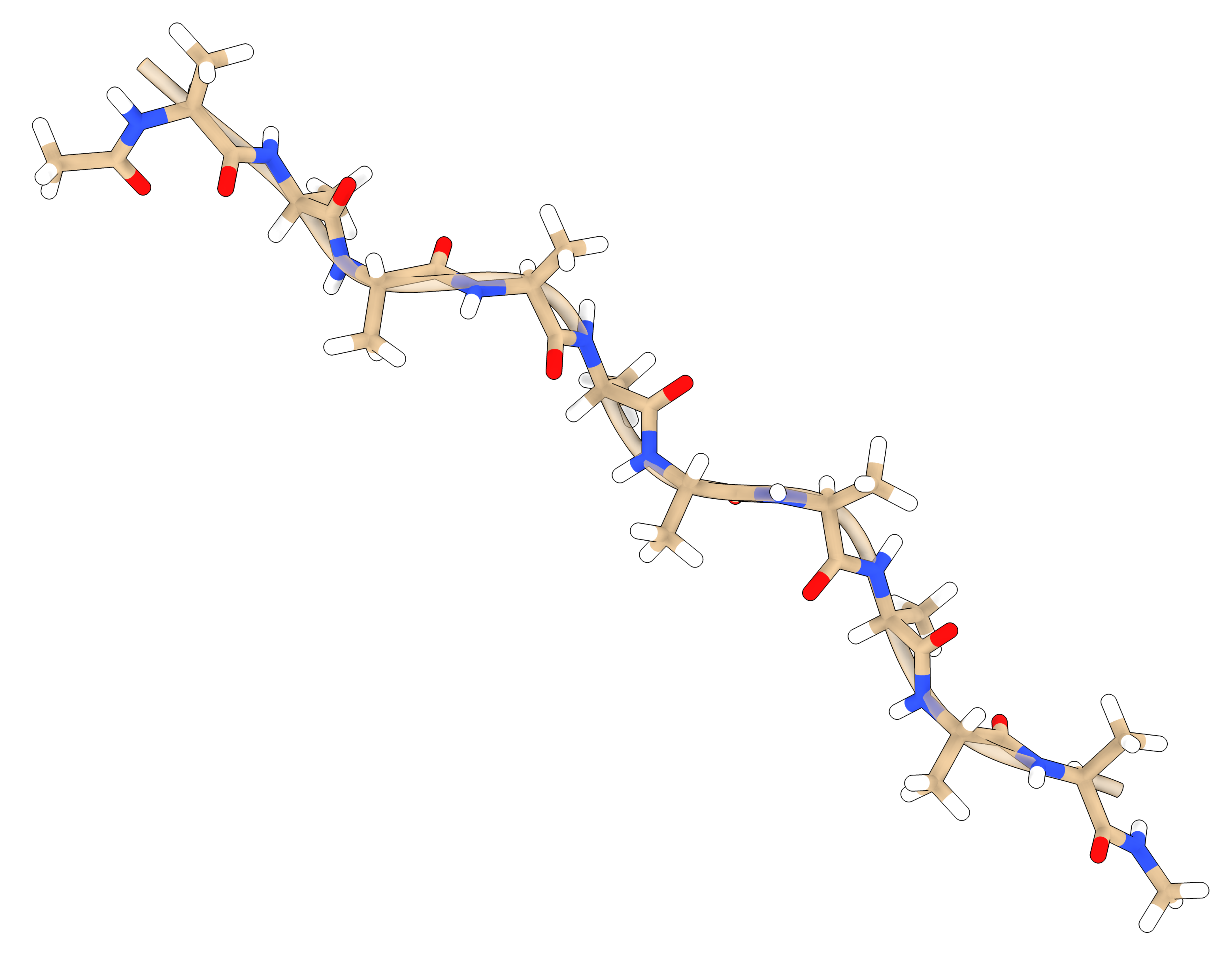} & High similarity for predictions & The poly-proline conformation is closely related to the $\beta$-sheet, but HNN is able to discern the two secondary structures \\
        \hline
        0 \includegraphics[width=0.08\textwidth]{images/deca_beta.pdf} & 8 \includegraphics[width=0.08\textwidth]{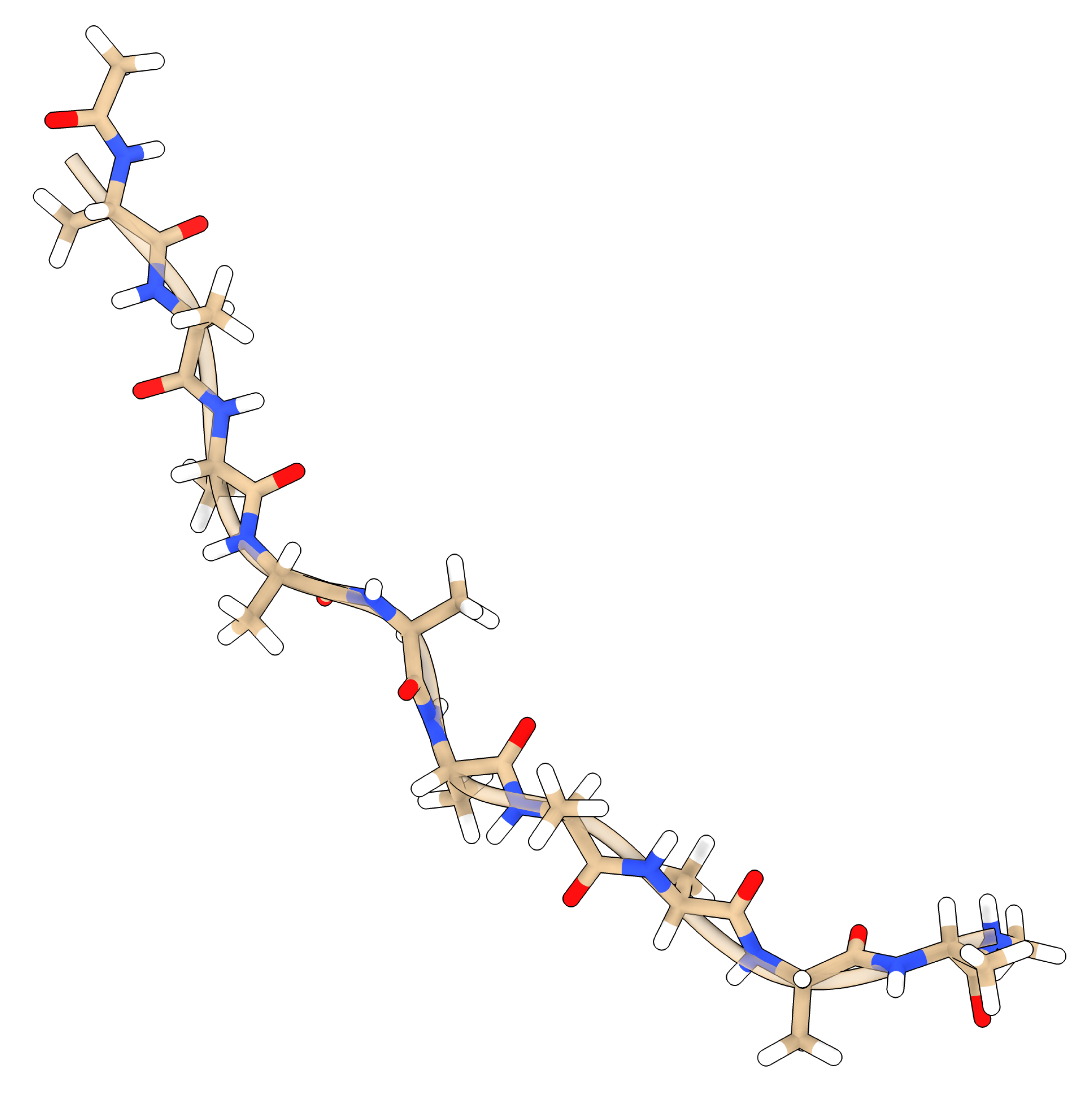} & High similarity for predictions & See explanation for clusters 0 and 3 \\ 
        \hline
        1 \includegraphics[width=0.09\textwidth]{images/deca_c1.pdf} & 2 \includegraphics[width=0.09\textwidth]{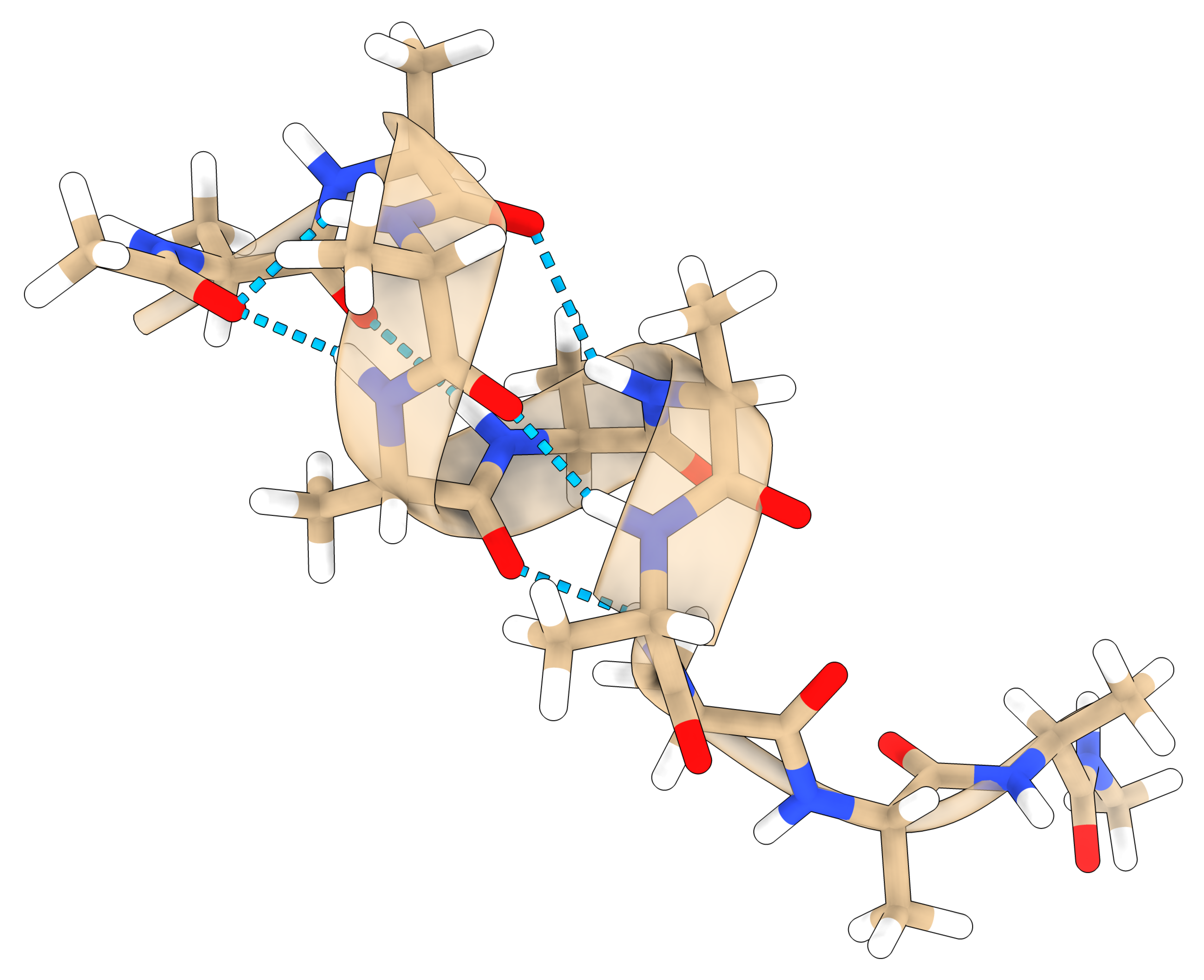} & High similarity for predictions & Cluster 2 is mainly organized in $\alpha$-helix, but part of it has unfolded structures, thus justifying the observed prediction differences \\
        \hline
        2 \includegraphics[width=0.09\textwidth]{images/deca_c2.pdf} & 4 \includegraphics[width=0.09\textwidth]{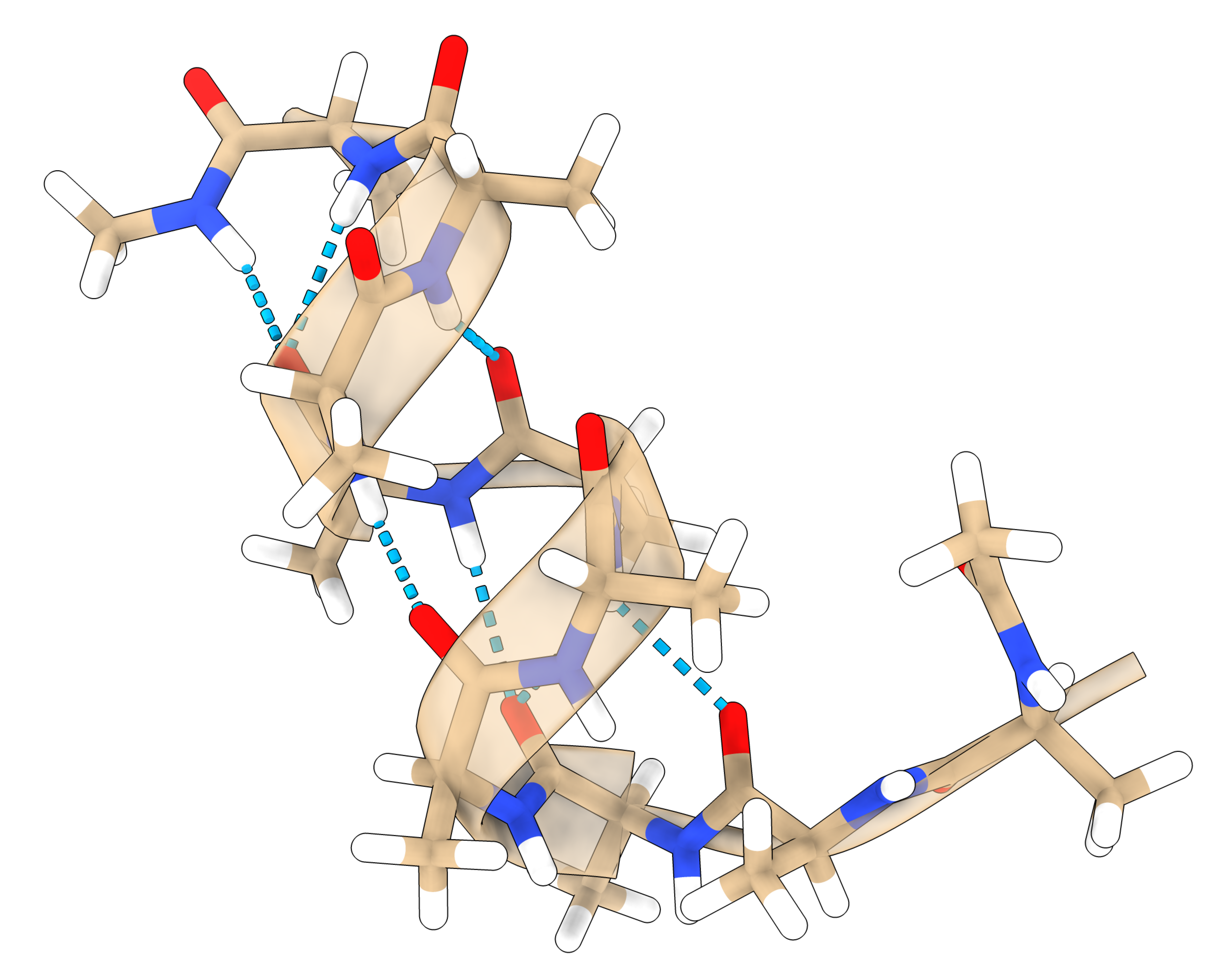} & High similarity for predictions & Cluster 2 and cluster 4 are similar both in terms of structures and distribution of dihedral angles, however they are considered different by HNN representing an outlier. See ``Secondary structure recognition'' section for discussion. \\
        \hline
        2 \includegraphics[width=0.09\textwidth]{images/deca_c2.pdf} & 9 \includegraphics[width=0.09\textwidth]{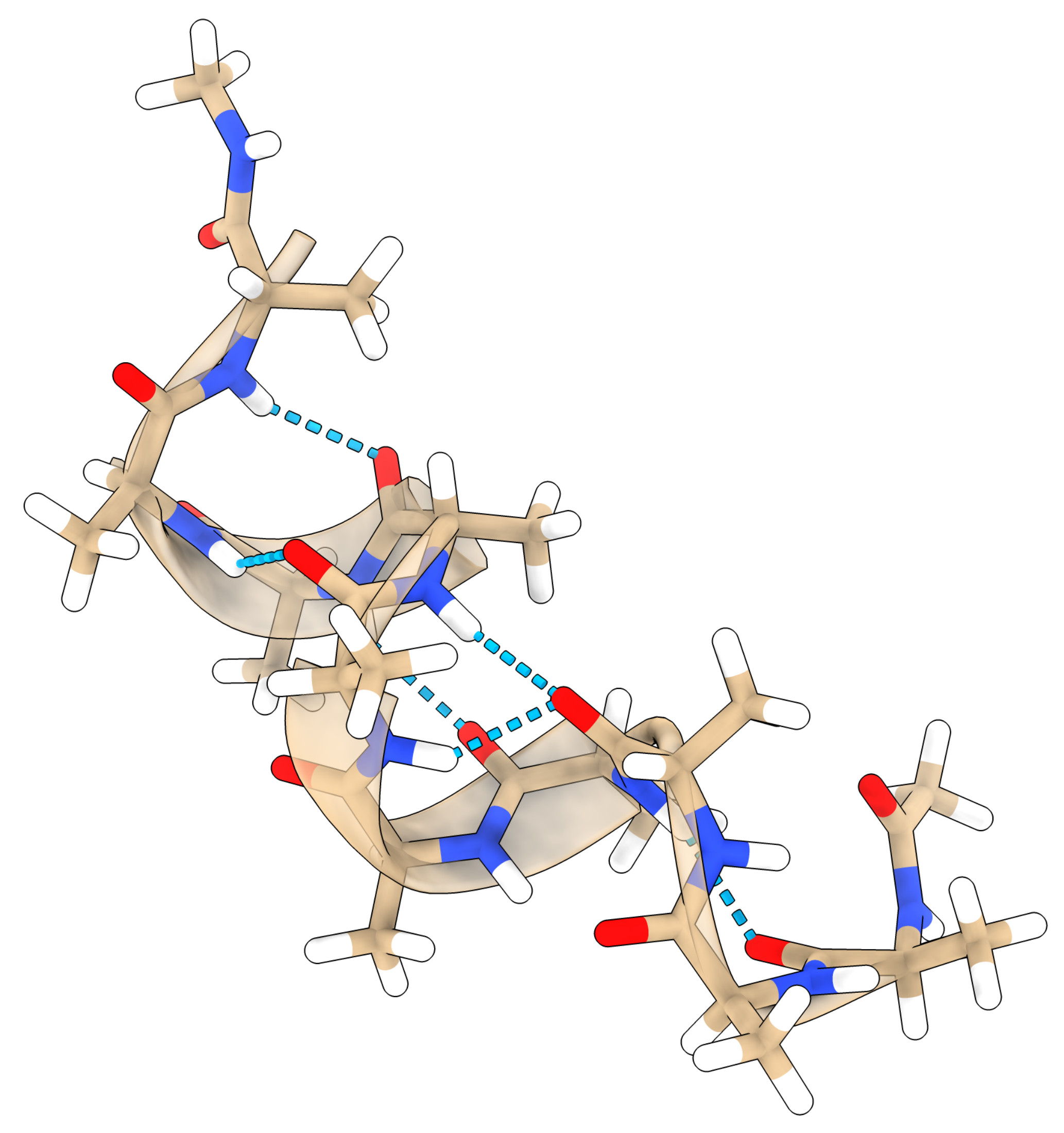} & High similarity for predictions & As for clusters 2 and 4, also 2 and 9 are structurally and energetically similar, but predicted different by HNN. See ``Secondary structure recognition'' section for discussion. \\
        \hline
        3 \includegraphics[width=0.09\textwidth]{images/deca_c3.pdf} & 6 \includegraphics[width=0.09\textwidth]{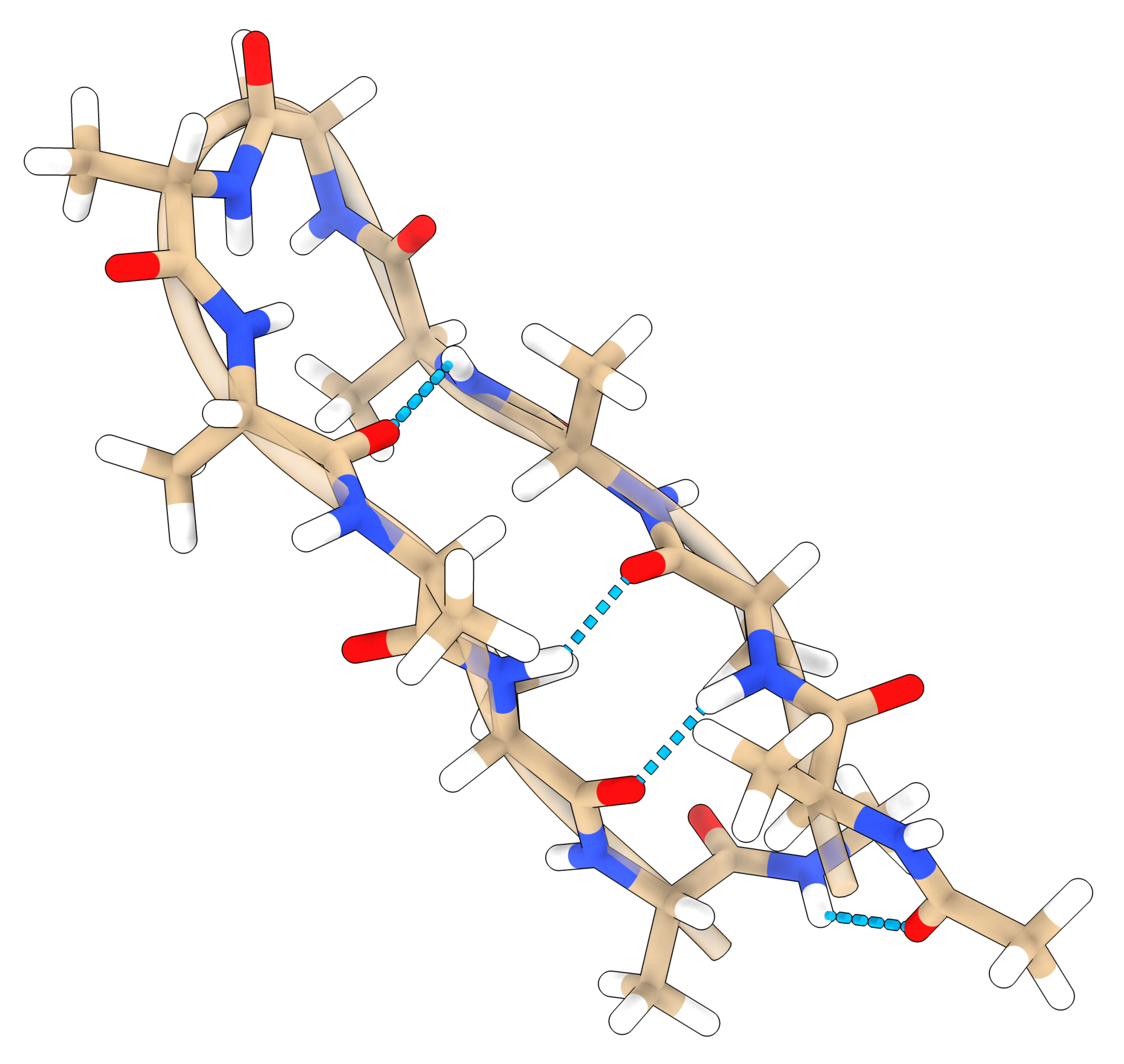} & Low similarity for predictions & Cluster 3 has part of its structure organized as $\beta$-sheet, similarly to cluster 6 \\
        \hline
        3 \includegraphics[width=0.09\textwidth]{images/deca_c3.pdf} & 7 \includegraphics[width=0.09\textwidth]{images/deca_c7.pdf} & High similarity for predictions & Cluster 3 does not have a perfect poly-proline structure, while cluster 7 does, thus justifying the differences in the predictions \\
        \hline
        7 \includegraphics[width=0.09\textwidth]{images/deca_c7.pdf} & 8 \includegraphics[width=0.08\textwidth]{images/deca_c8.pdf} & High similarity for predictions & See explanation for clusters 3 and 7  \\
        \hline
    \end{tabular}
\end{table}

\clearpage
\section{Discussion}
\label{sec:discussion}

Transfer learning provides a framework that allows making predictions on problems with limited available data starting from different, yet related datasets. Such a framework has been extensively used in various applications, however, so far in computational chemistry it has found few applications, mostly to approximate quantum-mechanical calculations or infer material and molecular properties \cite{smith2019approaching,cai2020,yamada2019}.
On the other hand, assessing free-energy values in conformational sampling by means of machine learning is, to the best of our knowledge, a novel and intriguing field of research that has recently seen more and more interest in the scientific community \cite{noe2020machine,bonati2021}.

In this paper, structural features of molecules are paired with free-energy estimates of a known molecular system in order to distinguish between high or low free-energy conformations of a target system whose free-energy surface is not known, and hence not used during training. This is accomplished by means of transfer learning, which allows to exploit the information gathered on a dataset to make predictions on a different one. The proposed methodology can be of great use since it would completely replace lengthy and expensive simulations, being substituted by a machine learning model that, once trained, can output free-energy estimates in a fraction of the time.
The proposed methodology, dubbed HNN in the paper, consists of two ingredients: (i) a novel hypergraph-based representation of molecules and (ii) a novel neural network model that can process hypergraphs as inputs and make decisions accordingly. More specifically, in this work we focused on a classification problem, aimed at classifying conformations in two classes, denoting high and low free-energy values.
The proposed hypergraph representation allows us to fully encode multi-atom interactions of a molecular system, since it describes the interactions between two, three, and four atoms. This innovative representation goes beyond well-known graph-based representations of molecules, that are limited to modeling pairwise interactions only.
The free-energy is then estimated for the target system using structural and free-energy data describing the smaller system through non-linear, black-box processing of information by means of the proposed neural network.
In this respect, our work represents one of the first of its kind with the use of hypergraphs for representing the chemico-physical properties of a given molecule, thus marking a significant advance in the field of machine learning and molecular simulations.

As a first case study, we considered the problem of classifying tri-alanine conformations starting from the information gathered from the smallest possible building block, i.e. alanine dipeptide. This first test was done to assess the capability of the proposed method in a controlled setting. In fact, the tri-alanine molecule is not big enough to populate organized secondary structures such as helices and $\beta$-sheets, and its three-dimensional structure can be seen as a combination of three different alanine dipeptides. The obtained results showed the ability of HNN to classify the tri-alanine conformations with a remarkably high AUC value.
Then, we moved to a more realistic molecular system, i.e. the deca-alanine system.
This experiment was considerably more challenging since the conformational properties of deca-alanine are significantly different from the data used during training, i.e. torsion angle values of the alanine backbone in low free-energy conformations are different between tri-alanine (used as training set) and deca-alanine.
Our results show that the HNN model successfully classifies low and high energy deca-alanine conformations with a remarkable degree of confidence, as showed in the Results section.

In addition to classifying low/high free-energy conformations in a supervised setting, we considered the application of the proposed methodology in an unsupervised setting. More precisely, we considered the possibility to cluster conformations of deca-alanine by using only structural information from alanine and tri-alanine, i.e. no free-energy values are used during training in this case.
Our results show that the HNN model is able to detect small conformational changes among all the analyzed clusters and to recognize similarities between conformations that belong to different cluster families. For instance, comparing the $p$-values computed for cluster 0 with respect to all the other clusters (see Tab. \ref{tab:deca-cluster-rel}), it is interesting to note that HNN predicts cluster 0 - corresponding to the fully extended conformation - energetically more similar to clusters 5 and 6 - forming a $\beta$-hairpin - rather than to other extended poly-proline like structures. Indeed, the $\beta$-hairpin is formed by two $\beta$-sheets in an antiparallel orientation connected by a turn that allows maximizing the number of intra-molecular h-bonds. Cluster 0 does not form a $\beta$-hairpin, but its backbone torsion angles assume values similar to those characterizing a $\beta$-sheet secondary structure that are detected by the model.
In general, the HNN model performs well in identifying clusters that are otherwise poorly classified by simple geometrical descriptors like RMSD (e.g., see the similarity between cluster 3 and 6).

The potential of such a model is huge. For instance, it might use simple building blocks (i.e., aminoacids) to predict low free-energy conformations of peptides, peptidoids -- often employed as drugs -- as well as of proteins or part of proteins not resolved by spectroscopic experiments. In this perspective, it is useful to better understand the model functionality with the aim of further improving its prediction capability. Examples are the differences predicted by the model for clusters 2-4 and 2-9, which are expected to be similar as both assume similar $\alpha$-helix conformations. The minor changes in the structural organization of the $\alpha$-helix between cluster 2, 4 and 9 suggest a remarkable sensitivity of the model in detecting such differences, however a deeper rationalization of the outlier data is necessary in the near future.
Our model has proven to be efficient in classifying low and high free-energy conformations in systems made by the same "building block" amino acid (alanine) in a relatively short sequence (deca-alanine). The capability of predicting free-energy values for more complex systems made by diverse secondary structures organized in tertiary structures and multiple amino acids, remains to be investigated. In this perspective, the results obtained for clusters 5, 6 and 0, where 5 and 6 form a tertiary structure not present in 0, are encouraging.

Furthermore, although the classification performance of the HNN model was satisfactory, as demonstrated by the remarkably high AUC, the performance of the model in a regression setting to predict conformations' free-energy values was not equally good (results not shown). The free-energy prediction in a regression setting is certainly a fascinating and desirable objective to pursue in the near future, since such information is particularly useful to elucidate molecular properties (e.g. to obtain an accurate description of the free-energy landscape) and design experiments accordingly \cite{limongelliligand,king2021}.

In conclusion, our work is a proof of concept that hypergraph-based neural networks can be successfully used to predict energetic properties for molecular systems that are otherwise inaccessible through state-of-the-art molecular simulations.
Our results prompt further work in this direction, notably on developing improved neural network models and hypergraph representations able to deal with even more complex, biologically relevant systems (e.g. protein-ligand complexes), marking a significant advance in the field of molecular simulations.
Finally, we note that the proposed methodology could be implemented as a run-time plug-in or a post-processing tool for molecular dynamics simulations, to identify low and high free-energy conformations that could help driving the sampling of the phase space, disclosing energetic and structural properties in an affordable computational time.

\section{Methods}
\label{sec:methods}

\subsection{Molecular representation in MD}
The starting data used by our methodology are structural and topological information coming from MD simulations. Therefore, in this section we will describe the main features about a molecule from the computational chemistry viewpoint. In MD simulations, a molecule is generally defined by a coordinate file, storing the Cartesian coordinates of each atom of the system, and a topology file, containing parameters to reproduce the physical properties of atoms.
For the sake of this study, we focus on the parameters that are relevant in the classification of conformational states for the various systems under study.
Understanding how a peptide or a protein is organized in the three-dimensional space in a given environment is not simple. Here, we evaluate if the information extracted from simulating a simple molecule could be used to classify the conformations of a more complex structure. As for the experiments reported in the Results section, we need to introduce two structural levels of peptides:
\begin{itemize}
    \item primary structure - the sequence containing a list of all the amino acids comprising a given peptide (e.g., ACE-ALA-ALA-ALA-NME for tri-alanine);
    \item secondary structure - the three-dimensional organization of all the residues in the sequence, which might give rise to well-known patterns, like helices, $\beta$-sheets, etc.
\end{itemize}

\begin{figure}[ht]
    \centering
    \includegraphics[scale=0.15]{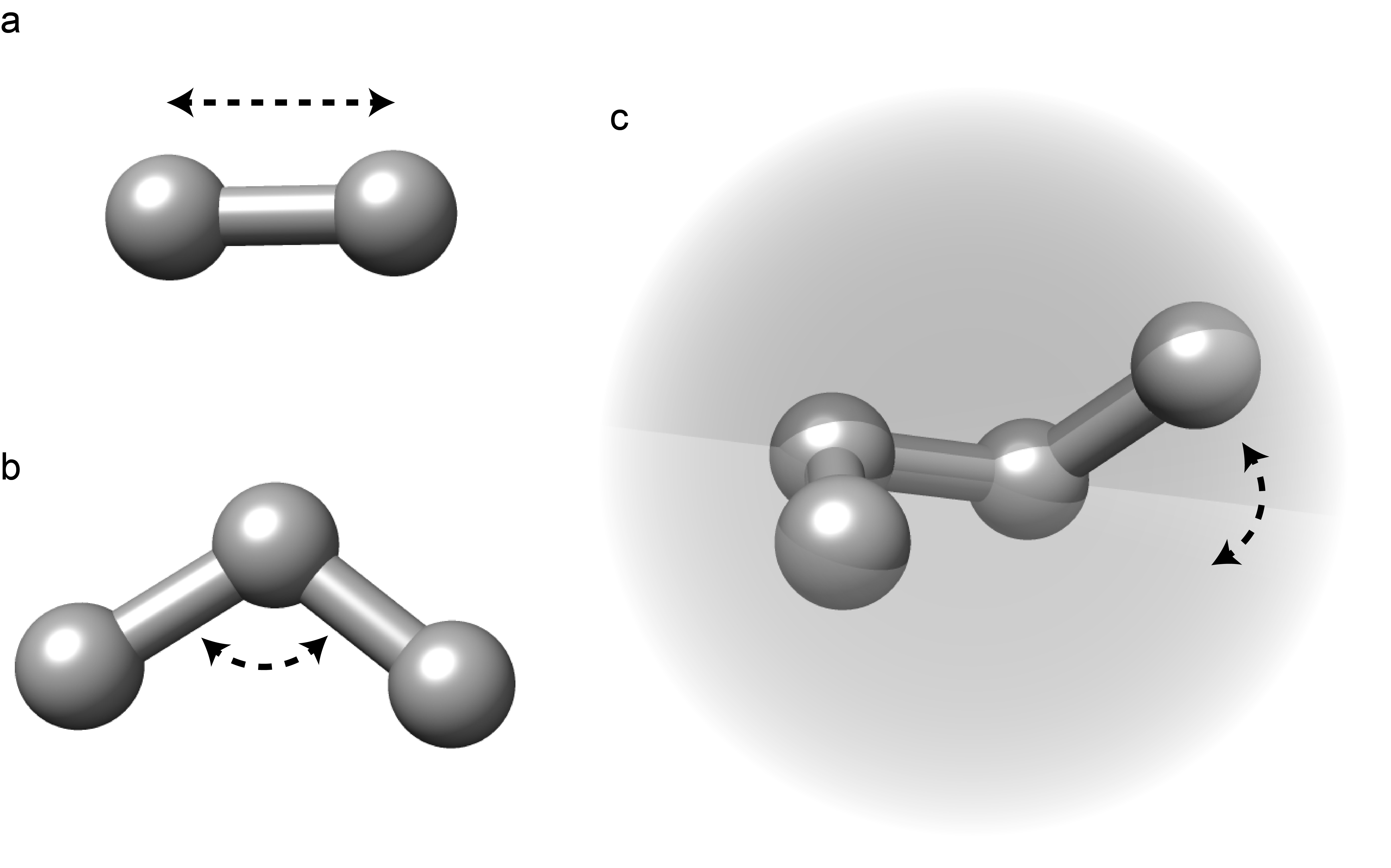}
    \caption{Graphical representation of the bond (a), angle (b), and dihedral angle (c) constraints in defining a molecule in MD simulations.}
    \label{bonded}
\end{figure}

For the latter, a key role is played by the values of the dihedral angles. Given four consecutive atoms (from 1 to 4) connected by bonds (i.e, 1 is bond to 2, 2 to 3, and 3 to 4), a dihedral (torsion) angle is the angle defined by two planes made by the first three atoms (1 to 3) and the second three (2 to 4). The result could be seen as a rotation around the bond between atoms 2 and 3, as represented in Fig. \ref{bonded}-c.

Each amino acid has two main dihedral angles running through its backbone structure that are called $\phi$ and $\psi$, and their combination can be mapped in order to assign known secondary structure motifs to a peptide, as it has been done in the Ramachandran plots of deca-alanine in Fig. \ref{fig:clusters}. Dihedral angles are included in the topology file (in fact, they contribute to the potential energy, as seen in Equation \ref{eq:potential_energy}) and they are provided as input to the HNN model during training. It is important to note that in order to define a specific conformation of a peptide by dihedral angles, the latter are defined only by consecutive atoms that are physically connected through bonds.

\subsection{Molecule representation as hypergraphs}

Due to the importance of higher-order interactions among atoms in describing the potential energy of conformations \eqref{eq:potential_energy}, we developed a novel hypergraph-based representation of molecules that encodes all relevant atom interactions.

Formally, a hypergraph $H(V,E,X,W)$ represents a conformation of a molecule, with $V$ being the set of all vertices (corresponding to the atoms of the molecule) and $E$ the set of hyperedges, modeling higher-order interactions, i.e. interactions between two or more vertices. Note that $E\subset\mathcal{P}(V)$, where $\mathcal{P}(V)$ is the power set of $V$, i.e. the set of all possible subsets of atoms.
We consider the following four interactions: bonds and non-bonds binary relations ($|e|=2$), angles ($|e|=3$), and dihedrals ($|e|=4$). $X$ is a matrix containing atom features, including atomic number and p-charge. The matrix $W\in \R^{|E|\times 5}$ contains features of the hyperedges.
Notably, the $i$th row of the matrix $W$, $W_i$, is a vector of size five containing the following information:
\begin{equation}
    W_i = \begin{bmatrix}
      1 \text{ if $e_i$ is a bond, 0 otherwise} \\
      \text{Coulomb force if $e_i$ denotes a Coulomb interaction, 0 otherwise}\\
      \text{Van der Waals force if $e_i$ denotes Van der Waals interaction, 0 otherwise}\\
      \text{angle between three atoms if $|e|=3$, 0 otherwise}\\
      \text{dihedral between four atoms if $|e|=4$, 0 otherwise}
    \end{bmatrix}
\end{equation}

The structure of an hypergraph is represented through two matrices: a binary incident matrix $B\in\R^{|E|\times|V|}$ 
\begin{equation}
    B_{ij} = \begin{cases}
    1 & v_j \in e_i \\
    0 & \text{otherwise}
    \end{cases}
\end{equation}
and an adjacency list $L\in\R^{|E|\times 2}$, such that $L_i = [i , j]$ indicates that $v_j \in e_i$. Both $B$ and $L$ encode the same type of information, but they are used in different ways to speed-up the computations.
Notably, the adjacency list is required for operations on GPU, and the incident matrix for operations running on CPU.

\subsection{Hypergraph message passing neural network}

We design a novel message passing neural network capable to process hypergraph-structured data. The proposed neural network model performs a series of message passing operations on the input hypergraph followed by pooling layers to calculate a function over the whole input hypergraph. Similarly to message passing schemes in graph neural networks \cite{gilmer2017neural}, the use of message passing operations allows us to significantly reduce the number of learnable parameters, which, in turn, decreases the bias and the required amount of data for training.

The proposed message passing layer for hypergraphs employs sigmoid activation functions and performs sum aggregation.
The nodes prepare a message through a linear function followed by a sigmoid activation that is sent to their hyperedges, which are then aggregated and combined with the hyperedge's features and sent back to the nodes. Finally, both the nodes and hyperedges update their internal representation.
These operations are formalized as follows:
\begin{align}
    \label{eq:hmpnn1}M_v &= f_v(X^{(t)}_v) \\
    \label{eq:hmpnn2}W^{(t+1)}_e &= g_w(W^{(t)}_e, \sum_{v \in e}M_v)\\
    \label{eq:hmpnn3}M_e &= f_w(W^{(t)}_e, \sum_{v \in e}M_v)\\
    \label{eq:hmpnn4}X^{(t+1)}_v &= g_v(X^{(t)}_v, \sum_{e \in e_v}M_e)
\end{align}
$X^{(t)}_v$ is the representation of vertex $v$ at layer $t$, $W^{(t)}_e$ is the representation of hyperedge $e$ at layer $t$, $f_v$ and $f_w$ are vertex and hyperedge messaging functions, respectively, both of which concatenate their inputs to form a vector and then apply sigmoidal functions on each element, $g_v$ and $g_w$ are vertex and hyperedge updating functions, respectively. The notation $e_v$ represents the set of hyperedges containing the vertex $v$. The updating functions apply a learnable linear transformation $L(x)$ to the current representation $x$ and add it to the incoming message $m$, i.e.,
\begin{align}
    g(x, m) &= L(x) + m\\
    L(x) &= Wx + b 
\end{align}
where $W$ and $b$ are the learnable parameters that are initialized randomly and then updated with back propagation during training.
The final output is the learned representation by the current layer.

\subsection{Pooling for hypergraphs}

After the message passing layers, a novel pooling layer is employed to produce a fixed size, numeric representation for the input hypergraph. This is achieved by comparing the hypergraph with a set of points of interest, which, for the specific application discussed in this study, are molecular conformations that have a distinct enough internal representation after message passing. The points of interest are selected via the kmeans clustering algorithm: cluster centroids are the points of interest.

Since the conformations of different molecular systems might have very different sizes, we need to devise a mechanism that allows us to make global decisions regardless of the system size. To this end, after the points of interest are computed, for each input hypergraph we create a fixed size feature vector encoding the hypergraph pair-wise similarity values with respect to the points of interest.
The similarity degree between an hypergraph $x$ and a point of interest $p$ is computed as follows. We consider the concatenation of all vertex and hyperedge features for both the hypergraph, denoted as $\mathbf{x}$, and the point of interest, denoted as $\mathbf{p}$. We note that $\mathbf{x}$ and $\mathbf{p}$ might have different sizes, and hence a direct comparisons is not possible. We therefore rely on a sliding window based mechanism that assesses their similarity by considering a sliding window with size equal to the smaller structure.

Formally, the pooling layer over input $\mathbf{x}$ with points of interest $P$, with $k=|P|$ defined by the user, performs the following steps:
\begin{itemize}
    \item For each point of interest $\mathbf{p}_i\in P$, create a vector $\mathbf{v}_i$ containing similarity values computed with the cosine similarity between $\mathbf{x}$ and $\mathbf{p}_i$ running over a sliding window with step 1
    \item Create vector $\mathbf{l}\in \R^{3|P|}$, and fill it in the following way:
    \begin{itemize}
        \item $\mathbf{l}_{3i} = \text{min}(\mathbf{v}_i)$
        \item $\mathbf{l}_{3i+1} = \text{average}(\mathbf{v}_i)$
        \item $\mathbf{l}_{3i+2} = \text{max}(\mathbf{v}_i)$
    \end{itemize}
    \item Feed $\mathbf{l}$ to a fully connected neural network, which outputs the probability that the input conformation $x$ is a low free-energy conformation
\end{itemize}

The learnable parameters in the proposed pooling mechanism are those of the final neural network. It is, however, important to update the representations of the points of interest periodically, e.g. when the message passing part of the network is updated.

\subsection{Scalability of the neural network operations}

The proposed molecule representation requires $3n^2 + 2n + 7e$ floating-point numbers per input conformation, where $n$ is the number of atoms and $e$ is the number of hyperedges. Each message passing layer of the neural network requires a constant amount of space to store the weights, hence it does not depend on the size of the molecules. However, it requires $O(e+n)$ time to perform the message passing operations, meaning that it scales linearly with respect to size of the molecule. The pooling layer, instead, requires $k \times (e+n)$ space and $O(k\times(e+n)^2)$ time, where $k$ is the number of interest points. Assuming $k$ is much smaller than $n$ and $e$, each pooling operation scales quadratically with the molecule size.

\subsection{Transfer learning}

Transfer learning \cite{weiss2016survey} is a machine learning technique used to learn models over some data distribution and transfer such models over a different distributions.
It is often described through its source and target distributions, as well as source and target tasks. The goal is to train a model to solve the source task on the source distribution, and then adjust it so that it can solve the target task on the target distribution.

In our experiments, we consider zero-shot transfer learning~\cite{xian2017zero} between a source and a target molecular system, e.g. between alanine dipeptide and tri-alanine. Zero-shot transfer learning does not assume the availability of information about the target system during training, making it more relevant in the molecular dynamics simulation setting we are interested in.
The task of interest is classification, and in particular we are interested in classifying low and high free-energy conformations.

To ensure that the message passing layers capture information relevant to the target system, we equip the loss function used during training with an extra regularization term in which input examples of the target system are partially processed by the network during training. Please note that such an extra regularization term does not take into account any supervised information we might have about the target system (i.e. its free-energy), but only structural information. To this end, we calculate a representative structure of the target system, and we pass it through the message passing layers.

Formally, for each target conformation observed during training, we construct a vector $D_i$ such that the $j$th entry is the $j$th feature of the related hypergraph. Stacking $D_i$ gives us a matrix, $D$. We calculate the principal axes of $D$ through their right singular vectors (eigenvectors of $A^TA$), and sum them, obtaining the representative $r_D$ for the target system. We note that $r_D$ represents an approximation of the variance of the target distribution.

The loss function used during training reads:
\begin{equation}
    \text{BinaryCrossEntropyLoss} + l_2 + \text{TargetLoss}
\end{equation}
where $l_2$ denotes the l2 penalty on the learnable weights, and the binary cross entropy loss is defined as
\begin{equation}
    \frac{1}{|N|}\sum_{i=1}^{N} y_i \log(p(y_i)) + (1-y_i) \log(1-p(y_i))
\end{equation}

The third term refers to the aforementioned extra regularization on the target system distribution:
\begin{equation}
\lVert\text{HMPNN}_2(\text{HMPNN}_1(r_D, W_1), W_2)\rVert_2
\end{equation}
where $\text{HMPNN}_i$ denotes the $i$-th message passing layer (without loosing generality, we assume two message passing layers, although this can be generalized to any number of layers) for hypergraphs with weights $W_i$, and $r_D$ is the representative defined as above.

\subsection{Unsupervised secondary structure recognition}
\label{sec:secondary_recognition}

Due to the lack of a ground truth for the deca-alanine free-energy landscape, we perform an additional test to validate the results of transfer learning. This test uses the trained HNN model to perform an unsupervised secondary structure recognition, assessing whether the HNN model is able to learn the secondary structures of the target system in a transfer learning setting.

We make the assumption that similar secondary structures of the target system have similar free-energy values, and that such similarities can be captured by relying only on the information of the source system used during training.
To test the validity of our assumption, we collect all predicted free-energy values for the structures in the various clusters, and compare their distributions with statistical tests to check for significant differences.
Notably, we used the Wilcoxon signed rank test \cite{rey2011wilcoxon} to check if the distributions underlying the prediction values are significantly different or not.
If the distributions are different according with a prescribed threshold ($p<0.01$), then we say that the HNN model predictions for the two clusters are in disagreement, i.e. they are significantly different.

\clearpage
\section*{Supplementary information}

\subsection*{Metadynamics simulations}
Metadynamics \cite{Laio2002} is an enhanced sampling technique that employs an external bias potential applied to one or more degrees of freedom (also known as collective variables - CVs), and constructed as a sum of Gaussian functions. Each single Gaussian is defined by the following expression:
\begin{equation}
    V(s,t) = \sum \omega e^{-\sum_{i=1}^d \frac{(s_i-s_i^0)^2}{2\sigma_i^2}}
\end{equation}
where $V$ is the total deposed bias, $\omega$ is the height of the Gaussian, $d$ is the number of CVs where the potential is deposed, $s$ is a given value in the CV, and $\sigma$ is the width of the Gaussian. 
The bias potential acts as an enhancer, moving the system out of any minimum encountered during the simulation: the deeper the minimum, the greater amount of potential will be placed on a given position $s$ along the CV, eventually overcoming all the barriers and entering in a semi-diffusive condition. Once the simulation has reached this level, and all possible states have been sampled, convergence is reached and the potential of mean force (PMF) can be reconstructed through the formula:
\begin{equation}
    V(s) = -F(s) + C
\end{equation}
where $F$ represents the free-energy along the chosen CVs and $C$ is a constant value.
A variant of the original metadynamics is the ``well-tempered'' (WT) \cite{Barducci2008} approach:
\begin{equation}
    V(s,t) = \sum \omega e^{\frac{-V(s,t)}{\Delta T}} e^{-\sum_{i=1}^d \frac{(s_i-s_i^0)^2}{2\sigma_i^2}}
\end{equation}
where an exponential term re-weights the height of the Gaussian based upon how much potential has already been placed on the same point $s$ at time $t$, and a parameter $\Delta$T which regulates how fast the Gaussian height decreases. In particular, $\Delta$T can also be seen as the difference in temperature between the hypothetical temperature felt by the enhanced CVs and the actual temperature of the simulation. To regulate this difference, the ``biasfactor'' is defined as $\gamma = (T + \Delta T)/T$, where $T$ is the temperature of the system. In such a way, convergence is reached faster and errors in free-energy estimates are dumped out. The WT approach might require an experienced user and the final free-energy estimate can be calculated through a slightly revised formula:
\begin{equation}
    V(s) = -\frac{\Delta T}{T + \Delta T} F(s) + C
\end{equation}

The input data for all three systems discussed in the present work have been obtained from well-tempered metadynamics simulations. All production runs were carried out in vacuum conditions and with the Amber FF14SB force field \cite{maier2015}. We used the Sander program of Amber18 together with PLUMED2 in order to activate the metadynamics algorithm \cite{case2018,tribello2014}. In the following, we list the settings employed for the three investigated different systems:
\begin{itemize}
    \item \textbf{alanine dipeptide}. This system has been thoroughly studied and it is known that the best CVs are the $\phi$ and $\psi$ dihedral angles of alanine. During the metadynamics simulations, $\sigma$ was set to 0.2 for $\phi$ and 0.3 for $\psi$, height was 1 kJ/mol, pace of 1000 steps, and biasfactor of 10;
    
    \item \textbf{tri-alanine}. We first sampled all possible structures using a general CV (i.e., RMSD of CA atoms), with a $\sigma$ of 0.007, a height of 1.5 kJ/mol, a pace of 500, and a biasfactor of 20. Then, we performed a Time-lagged Independent Component Analysis (TICA) \cite{noe2013,mccarty2017} using the three couples of $\phi$ and $\psi$ dihedrals in tri-alanine to construct two optimized CVs having the following construction:
    \begin{equation}
        CV1 = -0.0718 \  \phi 1 -0.0550 \  \psi 1 -0.9913 \  \phi 2 -0.0100 \  \psi 2 +0.0141 \  \phi 3 \  -0.0939 \  \psi 3
    \end{equation}
    \begin{equation}
        CV2 = -0.0524 \  \phi 1 -0.0931 \  \psi 1 +0.0894 \  \phi 2 -0.0973 \  \psi 2 +0.9719 \  \phi 3 -0.1630 \  \psi 3
    \end{equation}
    These two CVs are the two eigenvectors with the highest spectral gap among the six constructed by TICA, and they allowed us to discriminate 9 different conformational states, discussed in the main paper. The metadynamics parameter employed for the optimized run were 0.05 of $\sigma$ for both coordinates, 2 kJ/mol of height, pace of 500 steps, and biasfactor of 20;
    
    \item \textbf{deca-alanine}. We selected as CV the RMSD of the CA atoms of the residues, with a $\sigma$ of 0.007, a height of 0.1 kJ/mol, and a pace of 500 steps. 
\end{itemize}

\begin{figure}[ht]
    \centering
    \includegraphics[scale=0.75]{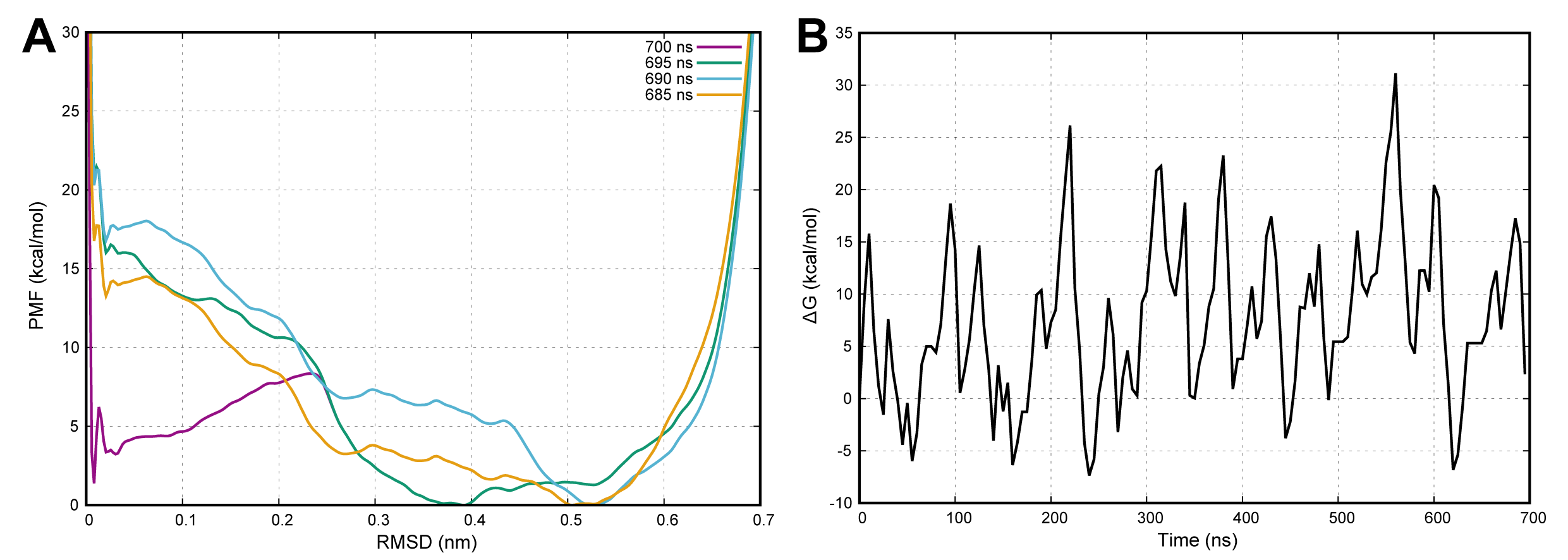}
    \caption{A. Super-imposed free-energy surfaces at different times for the Metadynamics simulation of deca-alanine. B. Plot of the free-energy difference with respect to time between the conformations close to 0 nm and those at around 0.5 nm of RMSD.}
    \label{fes}
\end{figure}

It is worth noting that different CVs settings have been used in the diverse systems investigated. As alanine dipeptide is a single aminoacid, the two backbone dihedral angles have been used as CVs, while for the tri-alanine peptide a linear combination of all of its dihedrals was employed. Instead, deca-alanine represents a much more complex system due to its higher structural complexity. Endowed with 20 dihedral angles, it is difficult to obtain from TICA a low number of CVs composed by a weighted linear combination of descriptors of the system. Such a difficulty might compromise the convergence of the free-energy calculation. For this reason, for deca-alanine we decided not to reach free-energy convergence, but to use a more general RMSD-based CV and the phase space sampling power of metadynamics to ensure the visit of the energetically most relevant states without computing the associated free-energy (Fig. \ref{fes}).

\subsection*{Molecular dynamics simulations}
Molecular Dynamics (MD) simulations were run for deca-alanine to sample structures around the ten different clusters that were obtained from metadynamics. The simulations were carried out in vacuum using the Amber FF14SB force field and the Sander algorithm of the Amber18 package \cite{maier2015,case2018}. During the simulations, we set a constraint on the RMSD of the deca-alanine CA atoms at 0.1 nm to obtain 1000 additional structures distributed around the representative one. Each run lasted for 1 ns with a time-step of 2 fs, for a total of 10000 structures that were used for the classification task.

\subsection*{Cluster creation in MD simulations}
For the deca-alanine system, a sample of structures were selected for the classification problem. These conformations were chosen by picking the ten most populated clusters over all the structures explored by the system during metadynamics. For this task, we employed the built-in functionality of Amber ``cluster'', which possesses several algorithms to associate a frame to a given family of structures \cite{case2018}. In particular, we opted for the ``hieragglo'' bottom-up default algorithm, with an $\epsilon$ of 2, applied only on the C$_\alpha$ atoms of the peptide. A total of 475 clusters were obtained, and the ten most populated were selected to be used for the secondary structure prediction test described in the main text.

\subsection*{Classification of low and high free-energy conformations}

To prepare training data for the model, we labelled any conformation with free-energy lower than 8 kJ/mol as low energy conformation, and everything else as high energy conformation.
The classification performed by the model operate as follows:
\begin{itemize}
    \item The model processes hypergraph representations of molecules and produces internal representations for the hypergraphs
    \item The internal representations are passed through a pooling layer, which assigns a fixed size vector representation to each hypergraph
    \item Each vector is inputted to a feed-forward neural network, assigning a probability $p\in[0, 1]$ of membership to the low-energy class
    \item Once all data are processed, we perform ROC analysis and compute the AUC, which gives us a robust measure of classification performance
\end{itemize}

\subsection*{$p$-values among clusters}

Details of the $p$-values for the comparisons shown in Table~\ref{tab:deca-cluster-rel}.

\definecolor{tblColorLThreshold}{rgb}{1,1,1}
\definecolor{tblColorLThresholdBad}{rgb}{1,1,0}
\definecolor{tblColorGThreshold}{rgb}{1,1,1}
\definecolor{tblColorGThresholdBad}{rgb}{1,1,0}

\begin{table}[ht!]
    \caption{$p$-values for the test assessing whether two clusters are in significant disagreement in terms of free-energy predictions. Cluster IDs are shown on row and column headings. Yellow cells indicate that the respective predictions' agreement was not inline with our original estimate based on the family of clusters.}
    \label{tab:deca-cluster-rel}
    \centering
    \begin{tabular}{|c||c|c|c|c|c|c|c|c|c|c|}\hline
     & 0 & 1 & 2 & 3 & 4 & 5 & 6 & 7 & 8 & 9    \\\hline\hline
0 & N/A & \cellcolor{tblColorLThreshold}1.96e-88 & \cellcolor{tblColorLThreshold}5.50e-55 & \cellcolor{tblColorLThresholdBad}3.63e-14 & \cellcolor{tblColorLThreshold}4.32e-77 & \cellcolor{tblColorLThreshold}1.02e-05 & \cellcolor{tblColorLThreshold}1.69e-09 & \cellcolor{tblColorLThresholdBad}1.89e-32 & \cellcolor{tblColorLThresholdBad}3.34e-18 & \cellcolor{tblColorLThreshold}2.51e-77 \\\hline
1 & \cellcolor{tblColorLThreshold}1.96e-88 & N/A & \cellcolor{tblColorLThresholdBad}7.80e-12 & \cellcolor{tblColorLThreshold}9.54e-54 & \cellcolor{tblColorGThreshold}0.0197 & \cellcolor{tblColorLThreshold}1.11e-66 & \cellcolor{tblColorLThreshold}1.67e-57 & \cellcolor{tblColorLThreshold}2.97e-37 & \cellcolor{tblColorLThreshold}3.71e-47 & \cellcolor{tblColorGThreshold}0.1582 \\\hline
2 & \cellcolor{tblColorLThreshold}5.50e-55 & \cellcolor{tblColorLThresholdBad}7.80e-12 & N/A & \cellcolor{tblColorLThreshold}2.18e-21 & \cellcolor{tblColorLThresholdBad}1.31e-07 & \cellcolor{tblColorLThreshold}2.04e-34 & \cellcolor{tblColorLThreshold}2.18e-28 & \cellcolor{tblColorLThreshold}2.65e-08 & \cellcolor{tblColorLThreshold}5.69e-16 & \cellcolor{tblColorLThresholdBad}1.55e-09 \\\hline
3 & \cellcolor{tblColorLThresholdBad}3.63e-14 & \cellcolor{tblColorLThreshold}9.54e-54 & \cellcolor{tblColorLThreshold}2.18e-21 & N/A & \cellcolor{tblColorLThreshold}8.60e-47 & \cellcolor{tblColorLThreshold}3.40e-05 & \cellcolor{tblColorGThresholdBad}0.0996 & \cellcolor{tblColorLThresholdBad}5.34e-07 & \cellcolor{tblColorGThreshold}0.1300 & \cellcolor{tblColorLThreshold}5.51e-46 \\\hline
4 & \cellcolor{tblColorLThreshold}4.32e-77 & \cellcolor{tblColorGThreshold}0.0197 & \cellcolor{tblColorLThresholdBad}1.31e-07 & \cellcolor{tblColorLThreshold}8.60e-47 & N/A & \cellcolor{tblColorLThreshold}1.86e-60 & \cellcolor{tblColorLThreshold}2.67e-49 & \cellcolor{tblColorLThreshold}7.06e-26 & \cellcolor{tblColorLThreshold}2.97e-37 & \cellcolor{tblColorGThreshold}0.4715 \\\hline
5 & \cellcolor{tblColorLThreshold}1.02e-05 & \cellcolor{tblColorLThreshold}1.11e-66 & \cellcolor{tblColorLThreshold}2.04e-34 & \cellcolor{tblColorLThreshold}3.40e-05 & \cellcolor{tblColorLThreshold}1.86e-60 & N/A & \cellcolor{tblColorGThreshold}0.0210 & \cellcolor{tblColorLThreshold}1.98e-17 & \cellcolor{tblColorLThreshold}1.97e-08 & \cellcolor{tblColorLThreshold}2.97e-64 \\\hline
6 & \cellcolor{tblColorLThreshold}1.69e-09 & \cellcolor{tblColorLThreshold}1.67e-57 & \cellcolor{tblColorLThreshold}2.18e-28 & \cellcolor{tblColorGThresholdBad}0.0996 & \cellcolor{tblColorLThreshold}2.67e-49 & \cellcolor{tblColorGThreshold}0.0210 & N/A & \cellcolor{tblColorLThreshold}7.36e-11 & \cellcolor{tblColorLThreshold}0.0012 & \cellcolor{tblColorLThreshold}5.91e-50 \\\hline
7 & \cellcolor{tblColorLThresholdBad}1.89e-32 & \cellcolor{tblColorLThreshold}2.97e-37 & \cellcolor{tblColorLThreshold}2.65e-08 & \cellcolor{tblColorLThresholdBad}5.34e-07 & \cellcolor{tblColorLThreshold}7.06e-26 & \cellcolor{tblColorLThreshold}1.98e-17 & \cellcolor{tblColorLThreshold}7.36e-11 & N/A & \cellcolor{tblColorLThresholdBad}0.0019 & \cellcolor{tblColorLThreshold}9.76e-27 \\\hline
8 & \cellcolor{tblColorLThresholdBad}3.34e-18 & \cellcolor{tblColorLThreshold}3.71e-47 & \cellcolor{tblColorLThreshold}5.69e-16 & \cellcolor{tblColorGThreshold}0.1300 & \cellcolor{tblColorLThreshold}2.97e-37 & \cellcolor{tblColorLThreshold}1.97e-08 & \cellcolor{tblColorLThreshold}0.0012 & \cellcolor{tblColorLThresholdBad}0.0019 & N/A & \cellcolor{tblColorLThreshold}3.04e-37 \\\hline
9 & \cellcolor{tblColorLThreshold}2.51e-77 & \cellcolor{tblColorGThreshold}0.1582 & \cellcolor{tblColorLThresholdBad}1.55e-09 & \cellcolor{tblColorLThreshold}5.51e-46 & \cellcolor{tblColorGThreshold}0.4715 & \cellcolor{tblColorLThreshold}2.97e-64 & \cellcolor{tblColorLThreshold}5.91e-50 & \cellcolor{tblColorLThreshold}9.76e-27 & \cellcolor{tblColorLThreshold}3.04e-37 & N/A \\\hline
    \end{tabular}
\end{table}

\clearpage
\bibliographystyle{abbrvnat}
\bibliography{bibliography.bib}

\begin{thebibliography}{66}
\providecommand{\natexlab}[1]{#1}
\providecommand{\url}[1]{\texttt{#1}}
\expandafter\ifx\csname urlstyle\endcsname\relax
  \providecommand{\doi}[1]{doi: #1}\else
  \providecommand{\doi}{doi: \begingroup \urlstyle{rm}\Url}\fi

\bibitem[Agostini et~al.(2012)Agostini, Vendruscolo, and
  Tartaglia]{Agostini2012237}
F.~Agostini, M.~Vendruscolo, and G.~G. Tartaglia.
\newblock {Sequence-Based Prediction of Protein Solubility}.
\newblock \emph{Journal of Molecular Biology}, 421\penalty0 (2-3):\penalty0
  237--241, 2012.
\newblock ISSN 0022-2836.
\newblock \doi{10.1016/j.jmb.2011.12.005}.

\bibitem[Bai et~al.(2021)Bai, Zhang, and Torr]{bai2021hypergraph}
S.~Bai, F.~Zhang, and P.~H. Torr.
\newblock Hypergraph convolution and hypergraph attention.
\newblock \emph{Pattern Recognition}, 110:\penalty0 107637, 2021.
\newblock \doi{10.1016/j.patcog.2020.107637}.

\bibitem[Barducci et~al.(2008)Barducci, Bussi, and Parrinello]{Barducci2008}
A.~Barducci, G.~Bussi, and M.~Parrinello.
\newblock Well-tempered metadynamics: A smoothly converging and tunable
  free-energy method.
\newblock \emph{Physical Review Letters}, 100\penalty0 (2), 2008.
\newblock \doi{10.1103/PhysRevLett.100.020603}.

\bibitem[Belkacemi et~al.(2022)Belkacemi, Gkeka, Lelievre, and
  Stoltz]{belkacemi2022}
Z.~Belkacemi, P.~Gkeka, T.~Lelievre, and G.~Stoltz.
\newblock Chasing collective variables using autoencoders and biased
  trajectories.
\newblock \emph{J. Chem. Theory Comput.}, 18:\penalty0 59--78, 2022.
\newblock \doi{10.1021/acs.jctc.1c00415}.

\bibitem[Bernardi et~al.(2015)Bernardi, Melo, and Scchulten]{bernardi2015}
R.~C. Bernardi, M.~C.~R. Melo, and K.~Scchulten.
\newblock Enhanced sampling techniques in molecular dynamics simulations of
  biological systems.
\newblock \emph{Biochim. Biophus. Acta}, 1850:\penalty0 872--877, 2015.
\newblock \doi{10.1016/j.bbagen.2014.10.019}.

\bibitem[Bodnar et~al.(2021)Bodnar, Frasca, Wang, Otter, Montufar, Li{\'o}, and
  Bronstein]{pmlr-v139-bodnar21a}
C.~Bodnar, F.~Frasca, Y.~Wang, N.~Otter, G.~F. Montufar, P.~Li{\'o}, and
  M.~Bronstein.
\newblock Weisfeiler and lehman go topological: Message passing simplicial
  networks.
\newblock In M.~Meila and T.~Zhang, editors, \emph{Proceedings of the 38th
  International Conference on Machine Learning}, volume 139 of
  \emph{Proceedings of Machine Learning Research}, pages 1026--1037. PMLR,
  18--24 Jul 2021.

\bibitem[Bonati et~al.(2021)Bonati, Piccini, and Parrinello]{bonati2021}
L.~Bonati, G.~Piccini, and M.~Parrinello.
\newblock Deep learning the slow modes for rare events sampling.
\newblock \emph{Proceedings of the National Academy of Sciences}, 118:\penalty0
  e2113533118, 2021.
\newblock \doi{10.1073/pnas.2113533118}.

\bibitem[B\=urkle et~al.(2021)B\=urkle, Perera, Gimbert, Nakamura, Kawata, and
  Asai]{burkle2021}
M.~B\=urkle, U.~Perera, F.~Gimbert, H.~Nakamura, M.~Kawata, and Y.~Asai.
\newblock Deep-learning approach to first-principles transport simulations.
\newblock \emph{Phys. Rev. Lett.}, 126:\penalty0 177701, 2021.
\newblock \doi{10.1103/PhysRevLett.126.177701}.

\bibitem[Butler et~al.(2018)Butler, Davies, Cartwright, Isayev, and
  Walsh]{butler2018machine}
K.~T. Butler, D.~W. Davies, H.~Cartwright, O.~Isayev, and A.~Walsh.
\newblock Machine learning for molecular and materials science.
\newblock \emph{Nature}, 559\penalty0 (7715):\penalty0 547--555, 2018.

\bibitem[Cai et~al.(2020)Cai, Wang, Xu, Zhang, Tang, Ouyang, Lai, and
  Pei]{cai2020}
C.~Cai, S.~Wang, Y.~Xu, W.~Zhang, K.~Tang, Q.~Ouyang, L.~Lai, and J.~Pei.
\newblock Transfer learning for drug discovery.
\newblock \emph{J. Med. Chem.}, 63:\penalty0 8683--8694, 2020.
\newblock \doi{10.1021/acs.jmedchem.9b02147}.

\bibitem[Case et~al.(2018)Case, Ben-Shalom, Brozell, Cerutti, {Cheatham III},
  Cruzeiro, Darden, Duke, Gilson, Gohlke, Goetz, Greene, Harris, Homeyer,
  Huang, Izadi, Kovalenko, Kurtzman, Lee, LeGrand, Li, Lin, Liu, Luchko, Luo,
  Mermelstein, Merz, Miao, Monard, Nguyen, Nguyen, Omelyan, Onufriev, Pan, Qi,
  Roe, Roitberg, Sagui, Schott-Verdugo, Shen, Simmerling, Smith, SalomonFerrer,
  Swails, Walker, Wang, Wei, Wolf, Wu, Xiao, York, and Kollman]{case2018}
D.~Case, I.~Ben-Shalom, S.~Brozell, D.~Cerutti, D.~{Cheatham III}, V.~Cruzeiro,
  T.~Darden, D.~Duke, R.E.;~Ghoreishi, M.~Gilson, H.~Gohlke, A.~Goetz,
  D.~Greene, R.~Harris, N.~Homeyer, Y.~Huang, S.~Izadi, A.~Kovalenko,
  T.~Kurtzman, T.~Lee, S.~LeGrand, P.~Li, C.~Lin, J.~Liu, T.~Luchko, R.~Luo,
  D.~Mermelstein, K.~Merz, Y.~Miao, G.~Monard, C.~Nguyen, H.~Nguyen,
  I.~Omelyan, A.~Onufriev, F.~Pan, R.~Qi, D.~Roe, A.~Roitberg, C.~Sagui,
  S.~Schott-Verdugo, J.~Shen, C.~Simmerling, J.~Smith, R.~SalomonFerrer,
  J.~Swails, R.~Walker, J.~Wang, H.~Wei, R.~Wolf, X.~Wu, L.~Xiao, D.~York, and
  P.~Kollman.
\newblock Amber 2018.
\newblock \emph{University of California, San Francisco}, 2018.

\bibitem[Ceriotti(2019)]{ceriotti2019unsupervised}
M.~Ceriotti.
\newblock Unsupervised machine learning in atomistic simulations, between
  predictions and understanding.
\newblock \emph{The Journal of chemical physics}, 150\penalty0 (15):\penalty0
  150901, 2019.

\bibitem[Chen et~al.(2019)Chen, Ye, Zuo, Zheng, and Ong]{chen2019graph}
C.~Chen, W.~Ye, Y.~Zuo, C.~Zheng, and S.~P. Ong.
\newblock Graph networks as a universal machine learning framework for
  molecules and crystals.
\newblock \emph{Chemistry of Materials}, 31\penalty0 (9):\penalty0 3564--3572,
  2019.
\newblock \doi{10.1021/acs.chemmater.9b01294}.

\bibitem[Chen et~al.(2022)Chen, Liu, Feng, Fu, Cai, Shao, and Chipot]{chen2022}
H.~Chen, H.~Liu, H.~Feng, H.~Fu, W.~Cai, X.~Shao, and C.~Chipot.
\newblock Mlcv: Bridging machine-learning-based dimensionality reduction and
  free-energy calculation.
\newblock \emph{J. Chem. Inf. Model.}, 62:\penalty0 1--8, 2022.
\newblock \doi{10.1021/acs.jcim.1c01010}.

\bibitem[Copeland(2016)]{copeland2016drug}
R.~A. Copeland.
\newblock The drug--target residence time model: a 10-year retrospective.
\newblock \emph{Nature Reviews Drug Discovery}, 15\penalty0 (2):\penalty0 87,
  2016.
\newblock \doi{10.1038/nrd.2015.18}.

\bibitem[Elton et~al.(2019)Elton, Boukouvalas, Fuge, and Chung]{elton2019deep}
D.~C. Elton, Z.~Boukouvalas, M.~D. Fuge, and P.~W. Chung.
\newblock Deep learning for molecular design--a review of the state of the art.
\newblock \emph{Molecular Systems Design \& Engineering}, 2019.
\newblock \doi{10.1039/C9ME00039A}.

\bibitem[Fawcett(2006)]{Fawcett_ROC}
T.~Fawcett.
\newblock {An Introduction to ROC Analysis}.
\newblock \emph{Pattern Recognition Letters}, 27\penalty0 (8):\penalty0
  861--874, June 2006.
\newblock ISSN 0167-8655.
\newblock \doi{10.1016/j.patrec.2005.10.010}.

\bibitem[Feng et~al.(2019)Feng, You, Zhang, Ji, and Gao]{feng2019hypergraph}
Y.~Feng, H.~You, Z.~Zhang, R.~Ji, and Y.~Gao.
\newblock Hypergraph neural networks.
\newblock In \emph{Proceedings of the AAAI Conference on Artificial
  Intelligence}, volume~33, pages 3558--3565, 2019.

\bibitem[Gilmer et~al.(2017)Gilmer, Schoenholz, Riley, Vinyals, and
  Dahl]{gilmer2017neural}
J.~Gilmer, S.~S. Schoenholz, P.~F. Riley, O.~Vinyals, and G.~E. Dahl.
\newblock Neural message passing for quantum chemistry.
\newblock In \emph{International conference on machine learning}, pages
  1263--1272. PMLR, 2017.

\bibitem[Hong et~al.(2021)Hong, Chun, Lee, Seo, Kang, and Han]{hong2021}
S.~J. Hong, H.~Chun, J.~Lee, M.~H. Seo, J.~Kang, and B.~Han.
\newblock First-principles-based machine-learning molecular dynamics for
  crystalline polymers with van der waals interactions.
\newblock \emph{J. Phys. Chem. Lett.}, 12\penalty0 (25):\penalty0 6000--6006,
  2021.
\newblock \doi{10.1021/acs.jpclett.1c01140}.

\bibitem[Jiang et~al.(2019)Jiang, Wei, Feng, Cao, and Gao]{jiang2019dynamic}
J.~Jiang, Y.~Wei, Y.~Feng, J.~Cao, and Y.~Gao.
\newblock Dynamic hypergraph neural networks.
\newblock In \emph{International Joint Conference on Artificial Intelligence},
  pages 2635--2641, 2019.

\bibitem[Jin et~al.(2019)Jin, Barzilay, and Jaakkola]{jin2019multi}
W.~Jin, R.~Barzilay, and T.~Jaakkola.
\newblock Multi-resolution autoregressive graph-to-graph translation for
  molecules.
\newblock \emph{arXiv preprint arXiv:1907.11223}, 2019.

\bibitem[Joshi and Deshmukh(2021)]{joshi2021}
S.~Y. Joshi and S.~A. Deshmukh.
\newblock A review of advancements in coarse-grained molecular dynamics
  simulations.
\newblock \emph{Molecular Simulation}, 47:\penalty0 786--803, 2021.
\newblock \doi{10.1080/08927022.2020.1828583}.

\bibitem[King et~al.(2021)King, Aitchison, Li, and Luo]{king2021}
E.~King, E.~Aitchison, H.~Li, and R.~Luo.
\newblock Recent developments in free energy calculations for drug discovery.
\newblock \emph{Front. Mol. Biosci.}, 8:\penalty0 712085, 2021.
\newblock \doi{10.3389/fmolb.2021.712085}.

\bibitem[Kmiecik et~al.(2016)Kmiecik, Gront, Kolinski, Wieteska, Dawid, and
  Kolinski]{kmiecik2016}
S.~Kmiecik, D.~Gront, M.~Kolinski, L.~Wieteska, A.~E. Dawid, and A.~Kolinski.
\newblock Coarse-grained protein models and their applications.
\newblock \emph{Chem. Rev.}, 116:\penalty0 7898--7936, 2016.
\newblock \doi{10.1021/acs.chemrev.6b00163}.

\bibitem[Kokubo et~al.(2011)Kokubo, Hu, and Pettitt]{kokubo2011}
H.~Kokubo, C.~Hu, and B.~Pettitt.
\newblock Peptide conformational preferences in osmolyte solutions: transfer
  free energies of deca-alanine.
\newblock \emph{J. Am. Chem. Soc.}, 133:\penalty0 1849--1858, 2011.
\newblock \doi{10.1021/ja1078128}.

\bibitem[Kukol(2015)]{kukol2015molecular}
A.~Kukol.
\newblock \emph{Molecular Modeling of Proteins}, volume 1215.
\newblock Springer, 2015.

\bibitem[Laio and Parrinello(2002)]{Laio2002}
A.~Laio and M.~Parrinello.
\newblock Escaping free energy minima.
\newblock \emph{Proceeding of the National Academy of Sciences}, 99:\penalty0
  12562--12566, 2002.
\newblock \doi{10.1073/pnas.202427399}.

\bibitem[{Lamim Ribeiro} and Tiwary(2018)]{lamim2018toward}
J.~M. {Lamim Ribeiro} and P.~Tiwary.
\newblock Toward achieving efficient and accurate ligand-protein unbinding with
  deep learning and molecular dynamics through {RAVE}.
\newblock \emph{Journal of Chemical Theory and Computation}, 15\penalty0
  (1):\penalty0 708--719, 2018.
\newblock \doi{10.1021/acs.jctc.8b00869}.

\bibitem[Leach(2001)]{andrew2001molecular}
A.~R. Leach.
\newblock Molecular modeling: Principles and applications.
\newblock \emph{Prentice Hall, London}, 2001.

\bibitem[Lee et~al.(2021)Lee, You, Lee, Li, and Kim]{lee2021}
D.~Lee, D.~You, D.~Lee, X.~Li, and S.~Kim.
\newblock Machine-learning-guided prediction models of critical temperature of
  cuprates.
\newblock \emph{J. Phys. Chem. Lett.}, 12\penalty0 (26):\penalty0 6211--6217,
  2021.
\newblock \doi{10.1021/acs.jpclett.1c01442}.

\bibitem[Lelimousin et~al.(2016)Lelimousin, Limongelli, and
  Sansom]{lelimousin2016}
M.~Lelimousin, V.~Limongelli, and M.~S.~P. Sansom.
\newblock Conformational changes in the epidermal growth factor receptor: Role
  of the transmembrane domain investigated by coarse-grained metadynamics free
  energy calculations.
\newblock \emph{J. Am. Chem. Soc.}, 138:\penalty0 10611--10622, 2016.
\newblock \doi{10.1021/jacs.6b05602}.

\bibitem[Limongelli(2020)]{limongelliligand}
V.~Limongelli.
\newblock Ligand binding free energy and kinetics calculation in 2020.
\newblock \emph{Wiley Interdisciplinary Reviews: Computational Molecular
  Science}, page e1455, 2020.
\newblock \doi{10.1002/wcms.1455}.

\bibitem[Livi et~al.(2016)Livi, Giuliani, and
  Sadeghian]{ecoli_graph_complexity}
L.~Livi, A.~Giuliani, and A.~Sadeghian.
\newblock Characterization of graphs for protein structure modeling and
  recognition of solubility.
\newblock \emph{Current Bioinformatics}, 11\penalty0 (1):\penalty0 106--114,
  Jan. 2016.
\newblock \doi{10.2174/1574893611666151109175216}.

\bibitem[Maier et~al.(2015)Maier, Martinez, Kasavajhala, Wickstrom, Hauser, and
  Simmerling]{maier2015}
J.~A. Maier, C.~Martinez, K.~Kasavajhala, L.~Wickstrom, K.~E. Hauser, and
  C.~Simmerling.
\newblock ff14sb: Improving the accuracy of protein side chain and backbone
  parameters from ff99sb.
\newblock \emph{J. Chem. Theory Comput.}, 11:\penalty0 3696--3713, 2015.
\newblock \doi{10.1021/acs.jctc.5b00255}.

\bibitem[McCarty and Parrinello(2017)]{mccarty2017}
J.~McCarty and M.~Parrinello.
\newblock A variational conformational dynamics approach to the selection of
  collective variables in metadynamics.
\newblock \emph{Journal of Chemical Physics}, 147:\penalty0 204109, 2017.
\newblock \doi{10.1063/1.4998598}.

\bibitem[Miller et~al.(2020)Miller, Geiger, Smidt, and
  No{\'e}]{miller2020relevance}
B.~K. Miller, M.~Geiger, T.~E. Smidt, and F.~No{\'e}.
\newblock Relevance of rotationally equivariant convolutions for predicting
  molecular properties.
\newblock \emph{arXiv preprint arXiv:2008.08461}, 2020.

\bibitem[Mori et~al.(2020)Mori, Okazaki, Mori, Kim, and Matubayasi]{mori2020}
Y.~Mori, K.~Okazaki, T.~Mori, K.~Kim, and N.~Matubayasi.
\newblock Learning reaction coordinates via cross-entropy minimization:
  application to alanine dipeptide.
\newblock \emph{J. Chem. Phys.}, 153:\penalty0 054115, 2020.
\newblock \doi{10.1063/5.0009066}.

\bibitem[Noe and Nuske(2013)]{noe2013}
F.~Noe and F.~Nuske.
\newblock A variational approach to modeling slow processes in stochastic
  dynamical systems.
\newblock \emph{Multiscale Model Simul.}, 11:\penalty0 635--655, 2013.
\newblock \doi{10.1137/110858616}.

\bibitem[No{\'e} et~al.(2020{\natexlab{a}})No{\'e}, De~Fabritiis, and
  Clementi]{noe2020machine}
F.~No{\'e}, G.~De~Fabritiis, and C.~Clementi.
\newblock Machine learning for protein folding and dynamics.
\newblock \emph{Current Opinion in Structural Biology}, 60:\penalty0 77--84,
  2020{\natexlab{a}}.
\newblock \doi{10.1016/j.sbi.2019.12.005}.

\bibitem[No{\'e} et~al.(2020{\natexlab{b}})No{\'e}, Tkatchenko, M{\"u}ller, and
  Clementi]{doi:10.1146/annurev-physchem-042018-052331}
F.~No{\'e}, A.~Tkatchenko, K.-R. M{\"u}ller, and C.~Clementi.
\newblock Machine learning for molecular simulation.
\newblock \emph{Annual Review of Physical Chemistry}, 71\penalty0 (1):\penalty0
  361--390, 2020{\natexlab{b}}.
\newblock \doi{10.1146/annurev-physchem-042018-052331}.

\bibitem[Ozer et~al.(2012)Ozer, Quirk, and Hernandez]{ozer2012}
G.~Ozer, S.~Quirk, and R.~Hernandez.
\newblock Thermodynamics of decaalanine stretching in water obtained by
  adaptive steered molecular dynamics simulations.
\newblock \emph{J. Chem. Theory Comput.}, 8:\penalty0 4837--4844, 2012.
\newblock \doi{10.1021/ct300709u}.

\bibitem[Ozer et~al.(2014)Ozer, Keyes, Quirk, and Hernandez]{ozer2014}
G.~Ozer, T.~Keyes, S.~Quirk, and R.~Hernandez.
\newblock Multiple branched adaptive steered molecular dynamics.
\newblock \emph{J. Chem. Phys.}, 141:\penalty0 064101, 2014.
\newblock \doi{10.1063/1.4891807}.

\bibitem[Palmer et~al.(2021)Palmer, Maasch, Torres, {de la Fuente-Nunez}, and
  Richardson]{palmer2021}
N.~Palmer, J.~R. M.~A. Maasch, M.~D.~T. Torres, C.~{de la Fuente-Nunez}, and
  A.~R. Richardson.
\newblock Molecular dynamics for antimicrobial peptide discovery.
\newblock \emph{Infection and Immunity}, 89\penalty0 (4):\penalty0 e00703--20,
  2021.
\newblock \doi{10.1128/IAI.00703-20}.

\bibitem[Pietrucci(2017)]{PIETRUCCI201732}
F.~Pietrucci.
\newblock Strategies for the exploration of free energy landscapes: Unity in
  diversity and challenges ahead.
\newblock \emph{Reviews in Physics}, 2:\penalty0 32 -- 45, 2017.
\newblock ISSN 2405-4283.
\newblock \doi{https://doi.org/10.1016/j.revip.2017.05.001}.
\newblock URL
  \url{http://www.sciencedirect.com/science/article/pii/S2405428317300059}.

\bibitem[Post et~al.(2019)Post, Wolf, and Stock]{post2019}
M.~Post, S.~Wolf, and G.~Stock.
\newblock Principal component analysis of nonequilibrium molecular dynamics
  simulations.
\newblock \emph{J. Chem. Phys.}, 150:\penalty0 204110, 2019.
\newblock \doi{10.1063/1.5089636}.

\bibitem[Raniolo and Limongelli(2020)]{raniolo2020}
S.~Raniolo and V.~Limongelli.
\newblock Ligand binding free-energy calculations with funnel metadynamics.
\newblock \emph{Nature Protocols}, 15:\penalty0 2837--2866, 2020.
\newblock \doi{10.1038/s41596-020-0342-4}.

\bibitem[Rey and Neuh{\"a}user(2011)]{rey2011wilcoxon}
D.~Rey and M.~Neuh{\"a}user.
\newblock Wilcoxon-signed-rank test.
\newblock In \emph{International Encyclopedia of Statistical Science}, pages
  1658--1659. Springer, 2011.

\bibitem[Sanchez-Lengeling and Aspuru-Guzik(2018)]{sanchez2018inverse}
B.~Sanchez-Lengeling and A.~Aspuru-Guzik.
\newblock Inverse molecular design using machine learning: {G}enerative models
  for matter engineering.
\newblock \emph{Science}, 361\penalty0 (6400):\penalty0 360--365, 2018.
\newblock \doi{10.1126/science.aat2663}.

\bibitem[Schmidt et~al.(2019)Schmidt, Marques, Botti, and
  Marques]{schmidt2019recent}
J.~Schmidt, M.~R.~G. Marques, S.~Botti, and M.~A.~L. Marques.
\newblock Recent advances and applications of machine learning in solid-state
  materials science.
\newblock \emph{npj Computational Materials}, 5\penalty0 (1):\penalty0 1--36,
  2019.
\newblock \doi{10.1038/s41524-019-0221-0}.

\bibitem[Schuetz et~al.(2017)Schuetz, {de Witte}, Wong, Knasmueller, Richter,
  Kokh, Sadiq, Bosma, Nederpelt, Heitman, et~al.]{schuetz2017kinetics}
D.~A. Schuetz, W.~E.~A. {de Witte}, Y.~C. Wong, B.~Knasmueller, L.~Richter,
  D.~B. Kokh, S.~K. Sadiq, R.~Bosma, I.~Nederpelt, L.~H. Heitman, et~al.
\newblock Kinetics for drug discovery: an industry-driven effort to target drug
  residence time.
\newblock \emph{Drug Discovery Today}, 22\penalty0 (6):\penalty0 896--911,
  2017.
\newblock \doi{10.1016/j.drudis.2017.02.002}.

\bibitem[Shahbabaei and Kim(2022)]{shahbabaei2022}
M.~Shahbabaei and D.~Kim.
\newblock Nanofluidics for gas separation applications: The molecular dynamics
  simulation perspective.
\newblock \emph{Separation \& Purification Reviews}, 51\penalty0 (2):\penalty0
  245--260, 2022.
\newblock \doi{10.1080/15422119.2021.1918720}.

\bibitem[Sheu et~al.(2003)Sheu, Yang, Selzle, and Schlag]{sheu2003}
S.~Sheu, D.~Yang, H.~Selzle, and E.~Schlag.
\newblock Energetics of hydrogen bonds in peptides.
\newblock \emph{Proc. Natl. Acad. Soc. USA}, 100:\penalty0 12683--12687, 2003.
\newblock \doi{10.1073/pnas.2133366100}.

\bibitem[Shukla and Tripathi(2021)]{shukla2021}
R.~Shukla and T.~Tripathi.
\newblock \emph{Molecular Dynamics Simulation in Drug Discovery: Opportunities
  and Challenges}, pages 295--316.
\newblock Springer Singapore, Singapore, 2021.
\newblock ISBN 978-981-15-8936-2.
\newblock \doi{10.1007/978-981-15-8936-2_12}.
\newblock URL \url{https://doi.org/10.1007/978-981-15-8936-2_12}.

\bibitem[Singh and Li(2019)]{singh2019}
N.~Singh and W.~Li.
\newblock Recent advances in coarse-grained models for biomolecules and their
  applications.
\newblock \emph{Int. J. Mol. Sci.}, 20:\penalty0 3774, 2019.
\newblock \doi{10.3390/ijms20153774}.

\bibitem[Smith et~al.(2019)Smith, Nebgen, Zubatyuk, Lubbers, Devereux, Barros,
  Tretiak, Isayev, and Roitberg]{smith2019approaching}
J.~S. Smith, B.~T. Nebgen, R.~Zubatyuk, N.~Lubbers, C.~Devereux, K.~Barros,
  S.~Tretiak, O.~Isayev, and A.~E. Roitberg.
\newblock Approaching coupled cluster accuracy with a general-purpose neural
  network potential through transfer learning.
\newblock \emph{Nature Communications}, 10\penalty0 (1):\penalty0 1--8, 2019.
\newblock \doi{10.1038/s41467-019-10827-4}.

\bibitem[Sultan and Pande(2018)]{sultan2018}
M.~Sultan and V.~Pande.
\newblock Automated design of collective variables using supervised machine
  learning.
\newblock \emph{J. Chem. Phys.}, 149:\penalty0 094106, 2018.
\newblock \doi{10.1063/1.5029972}.

\bibitem[Tiwary et~al.(2015)Tiwary, Limongelli, Salvalaglio, and
  Parrinello]{tiwary2015kinetics}
P.~Tiwary, V.~Limongelli, M.~Salvalaglio, and M.~Parrinello.
\newblock Kinetics of protein--ligand unbinding: Predicting pathways, rates,
  and rate-limiting steps.
\newblock \emph{Proceedings of the National Academy of Sciences}, 112\penalty0
  (5):\penalty0 E386--E391, 2015.
\newblock \doi{10.1073/pnas.1424461112}.

\bibitem[Tonge(2017)]{tonge2017drug}
P.~J. Tonge.
\newblock Drug--target kinetics in drug discovery.
\newblock \emph{ACS Chemical Neuroscience}, 9\penalty0 (1):\penalty0 29--39,
  2017.
\newblock \doi{10.1021/acschemneuro.7b00185}.

\bibitem[Torrey and Shavlik(2010)]{torrey2010transfer}
L.~Torrey and J.~Shavlik.
\newblock Transfer learning.
\newblock In \emph{Handbook of research on machine learning applications and
  trends: algorithms, methods, and techniques}, pages 242--264. IGI global,
  2010.

\bibitem[Tribello et~al.(2014)Tribello, Bonomi, Branduardi, Camilloni, and
  Bussi]{tribello2014}
G.~A. Tribello, M.~Bonomi, D.~Branduardi, C.~Camilloni, and G.~Bussi.
\newblock Plumed 2: new feathers for an old bird.
\newblock \emph{Comp. Phys. Comm.}, 185:\penalty0 604--613, 2014.
\newblock \doi{10.1016/j.cpc.2013.09.018}.

\bibitem[Valsson et~al.(2016)Valsson, Tiwary, and
  Parrinello]{valsson2016enhancing}
O.~Valsson, P.~Tiwary, and M.~Parrinello.
\newblock Enhancing important fluctuations: Rare events and metadynamics from a
  conceptual viewpoint.
\newblock \emph{Annual Review of Physical Chemistry}, 67:\penalty0 159--184,
  2016.
\newblock \doi{10.1146/annurev-physchem-040215-112229}.

\bibitem[Weiss et~al.(2016)Weiss, Khoshgoftaar, and Wang]{weiss2016survey}
K.~Weiss, T.~M. Khoshgoftaar, and D.~Wang.
\newblock A survey of transfer learning.
\newblock \emph{Journal of Big data}, 3\penalty0 (1):\penalty0 1--40, 2016.
\newblock \doi{10.1186/s40537-016-0043-6}.

\bibitem[Xia et~al.(2021)Xia, Yin, Yu, Wang, Cui, and Zhang]{xia2021self}
X.~Xia, H.~Yin, J.~Yu, Q.~Wang, L.~Cui, and X.~Zhang.
\newblock Self-supervised hypergraph convolutional networks for session-based
  recommendation.
\newblock In \emph{Proceedings of the AAAI Conference on Artificial
  Intelligence}, volume~35, pages 4503--4511, 2021.

\bibitem[Xian et~al.(2017)Xian, Schiele, and Akata]{xian2017zero}
Y.~Xian, B.~Schiele, and Z.~Akata.
\newblock Zero-shot learning-the good, the bad and the ugly.
\newblock In \emph{Proceedings of the IEEE Conference on Computer Vision and
  Pattern Recognition}, pages 4582--4591, 2017.

\bibitem[Yamada et~al.(2019)Yamada, Liu, Wu, Koyama, Ju, Shiomi, Morikawa, and
  Yoshida]{yamada2019}
H.~Yamada, C.~Liu, S.~Wu, Y.~Koyama, S.~Ju, J.~Shiomi, J.~Morikawa, and
  R.~Yoshida.
\newblock Predicting materials properties with little data using shotgun
  transfer learning.
\newblock \emph{ACS Cent. Sci.}, 5:\penalty0 1717--1730, 2019.
\newblock \doi{10.1021/acscentsci.9b00804}.

\end{thebibliography}

\end{document}